\documentclass[a4paper,11pt]{article}
\usepackage[width=15cm,height=23cm]{geometry}
\usepackage[usenames]{color}
\usepackage{amsfonts,amssymb}
\usepackage{theorem}
\usepackage[dvips]{graphicx}

\long\def\symbolfootnote[#1]#2{\begingroup%
\def\thefootnote{\fnsymbol{footnote}}\footnote[#1]{#2}\endgroup}

\def\slshp{p\!\!\!\slash}

\def\slshpart{{\partial} \! \! \! \slash}
\def\slshW{W \! \! \! \! \! \slash}
\def\slshZ{Z  \! \! \! \! \slash}

\def \KL{K\"all\'en-Lehmann\,}

\begin{document}


\normalsize
\rightline{IFUM-996-FT}
\rightline{FR-PHENO-2012-11}

\vskip 2.0 truecm
\Large
\bf
\centerline{Heavy top renormalon contribution to fermion propagators}
\normalsize \rm

\large
\rm
\vskip 1.3 truecm
\centerline{D.~Bettinelli$^a$\footnote{e-mail: {\tt daniele.bettinelli@mi.infn.it}}, 
J.~J.~van der Bij$^b$\footnote{e-mail: {\tt jochum@physik.uni-freiburg.de}}}

\normalsize
\medskip
\begin{center}
$^a$INFN, Sezione di Milano\\
via Celoria 16, I-20133 Milano, Italy\\
$^b$Physikalisches Institut, Albert-Ludwigs-Universit\"at Freiburg\\
Hermann-Herder-Str. 3, D-79104 Freiburg im Breisgau, Germany.
\end{center}

\vskip 0.7  truecm
\normalsize
\bf
\centerline{Abstract}

\rm

\begin{quotation}

We study resummed perturbative contributions due to a heavy top-quark.
These renormalon contributions are evaluated for fermion propagators.
Results for the top-quark width are given.
Estimates of non-perturbative uncertainties are made on the $\rho$-parameter
using different schemes of dealing with the Landau-pole.
For the physical top-quark mass the effects are negligible.

\end{quotation}

\newpage

\section{Introduction}
\label{sect.flavour}
%

In the electroweak sector of the Standard Model (SM) every particle acquires its mass through an interaction 
 with a scalar potential in a non-trivial vacuum. As a consequence, all the masses are proportional to a 
common scale, namely $G_F^{-1/2}$, which is fixed by low-energy measurements such as the $\mu$-decay rate.
In this situation the decoupling theorem \cite{Appelquist:1974tg} does not hold and thus there exist 
 low-energy observables in which the quantum effects induced by virtual heavy particles do not vanish when the mass 
of these particles goes to infinity.

Most prominent among the non-decoupling effects is the $\rho$-parameter \cite{Veltman:1977kh} 
 which provides a measure of the relative strength of neutral and charged current interactions 
 in four fermion processes at zero momentum transfer.
At tree level $\rho = 1$ due to a global accidental $SU(2)$ symmetry, the so-called custodial symmetry.
 $\rho$ can receive radiative corrections only by those sectors of the SM that break explicitly the custodial symmetry,
 namely the hypercharge and the Yukawa couplings that give  different masses to the components of fermion doublets.
 In the latter case the contribution to the $\rho$-parameter is proportional to the mass splitting, therefore 
 the leading contribution comes from the top-bottom doublet.

At one loop  the $\rho$-parameter has a quadratic dependence on the top-quark mass, 
$\Delta \rho^{(1)} \approx G_F\, m^2_t$, and a logarithmic dependence on the Higgs mass,
 $\Delta \rho^{(1)} \approx g'^{2}\, \log\Big(\frac{m_H}{M_W}\Big).$
Two-loop corrections at the leading order, i.e. $\Delta \rho^{(2)} \approx G_F^2\, m^4_t$, and at the next-to-leading
 order, i.e. $\Delta \rho^{(2)} \approx G^2_F\, m^2_t\, M^2_Z$, in the top-quark mass were computed in the limits 
 $m_H \to 0$ and $m_H \gg m_t$ in Refs.\cite{vanderBij:1986hy, Barbieri} and for arbitrary Higgs mass in 
Ref.\cite{Fleischer:1993ub}. It turned out that  due to accidental cancellations, the subleading corrections at two 
 loops are larger than the leading ones \cite{Vicini}.  At three loops the computation of the leading 
top-quark corrections, $\Delta \rho^{(3)} \approx G_F^3\,m^6_t$, 
in the massless Higgs limit,  was carried out in Ref.\cite{vanderBij:2000cg}.
The complete dependence on the Higgs mass at three loops was obtained in Ref.\cite{Faisst:2003px}. 
Numerically it was found that this contribution to $\Delta \rho$ is quite large and 
 provides a sizable correction ($\approx 36 \%$) to the leading electroweak correction at two loops.  
 However, the size of the three loop correction is only about $2 \%$ of the much larger two-loop subleading 
 electroweak correction. Moreover, the perturbative series of the leading top-quark contributions to the $\rho$-parameter 
has alternating signs up to three loops.

 This raises the issue of the convergence of the perturbative expansion (it might be that 
 this series is divergent, but Borel summable) and calls for a better 
understanding of higher order radiative corrections.
 It would be highly desirable to have a simplified framework in which the leading top-quark contributions to the 
 $\rho$-parameter can be computed to all orders in perturbation theory and eventually summed up.
The actual calculation of the leading radiative corrections in the top-quark mass 
is greatly simplified by the observation that to obtain them it is enough to consider
 the lagrangian of the SM in the 
 limit of vanishing gauge coupling constants $g, g' \to 0$ \cite{Barbieri}. 
 This gaugeless limit provides an efficient way of  reducing the number of Feynman diagrams to be computed 
 and it has been used in the two and three loop computations mentioned above. 

In some recent papers \cite{Bettinelli:2010gm,Bettinelli:2011zd} the effects of a finite top-width
were resummed by using a
 $SU(N_F)\times U(1)$ electroweak model in the large $N_F$-limit \cite{Einhorn:1984mr, Aoki:1990mb}.
In this paper we will use another resummation procedure.
 In the model we are going to study the symmetry group is the one 
of the SM, namely $SU(2) \times U(1)$, but instead of having three generations of quarks and leptons we consider 
 a large number of copies ($N_G$) of the third family of quarks. 
In order to avoid the presence of chiral anomalies we have to take 
into account an equal number of copies of the third family of leptons, however this will
play no role in the further calculations.
All of the extra quark doublets contain a massive particle, the top-quark with mass $m_t$, and 
a massless particle, the bottom-quark, while both components of  the extra lepton  doublets are 
taken to be massless.
Notice that this is not meant to be phenomenologically relevant. Indeed, from LEP ($Z \to \bar{\nu}\, \nu$)  
we know that there are only three generations of light neutrinos.
We take the approximation that  the CKM matrix is diagonal.
The large $N_G$-limit is performed by keeping $y_t^2\,N_G$ fixed, where $y_t$ is the top-Yukawa coupling.
In this limit only the graphs with a maximal number of fermion
loops contribute. This sort of sum is known as a renormalon chain \cite{Beneke:1998ui}.

As we are working in the limit of a heavy top-quark the effects of the gauge couplings can be ignored 
and we have a resummed propagator in the Higgs and the Goldstone-boson sector only.
The resulting Dyson propagators contain, in addition to the physical pole, a
 tachyon-pole in the euclidean region, $p^2 = -\Lambda^2_T$, 
which spoils causality and makes  the Wick-rotated Feynman integrals ill-defined.

If one wants to use the resummed propagators in further loop insertions one has to find a way to treat 
this tachyon-pole.
 In this connection the introduction of an UV cutoff 
 at $\Lambda < \Lambda_T$ has been proposed in Ref.\cite{Aoki:1992db}. However this procedure breaks 
 gauge invariance. 
We have adopted another strategy that was used successfully in Ref.\cite{Binoth:1997pd}. 
Assuming that the 
 occurrence of the tachyon-pole is not due to the inconsistency of the theory under consideration, 
but of the intermediary expansion technique used,  it is reasonable to simply subtract the tachyon-pole minimally
from the propagator, thereby restoring causality.
This is actually a rather old idea \cite{felmanetc}, that has been adapted in a slightly modified form
in QCD under the name of analytic perturbation theory \cite{shirkov}.

 One should be careful in doing this because the tachyon-pole 
contributes to the \KL spectral function. Further corrections might be needed in order to preserve
fundamental aspects of the theory. In analytic perturbation theory, for instance, the 
tachyon subtraction was done at the level of the effective charge. On the propagator level this
corresponds to subtracting the tachyon-pole and adding it back with the same strength 
at $p^2=0$. This is necessary in order to preserve asymptotic freedom.
Technically one deals with a subtracted dispersion relation.
In another context \cite{akhoury}, resummation inside the Higgs-propagator, the normalization of the 
spectral density was essential and one had to multiply the propagator with a constant 
non-perturbative factor.

Unfortunately the addition of non-perturbative factors is not unique, as was already mentioned
in the earliest paper \cite{felmanetc}. Nonetheless, it is important to get some idea
on the size of possible non-perturbative effects. A theory that is only defined in 
 the perturbative approximation is of course not satisfactory. This is also true for
the Standard Model. Ultimately, one will try to put the theory on the lattice
in order to go beyond perturbation theory. Since the electroweak sector of the SM is not asymptotically free, presumably
cut-off effects stay present in the lattice predictions. 
The situation is complicated due to the presence of fermion-doubles on the lattice,
which one cannot remove as easily as in lattice QCD by moving their mass to infinity, since they get their
mass via the Higgs mechanism and therefore become strongly coupled in this limit.
 In order to compare with the continuum
the use of resummed propagators is at the moment the only alternative, whereby the uncertainty 
due to the non-perturbative effects should correspond to   the uncertainty in predictions
due to the cut-off effects on the lattice. It is to be remarked however, that even with $m_t=172 GeV$ 
perturbation theory is quite satisfactory. As the Higgs boson is also presumably light, from the practical point
of view perturbation theory should be good enough for the SM.

In this paper we calculate the contribution of the resummed propagators to the
top and bottom-quark propagators. These contributions can then be used as input
for further calculations, but are of interest by themselves.
The outline of the paper is as follows.
In section~\ref{sect.self-en} we discuss the resummed Higgs and would-be Goldstone-bosons propagators.
In section~\ref{sect.rho}  we discuss possible non-perturbative contributions to the $\rho$-parameter
due to alternative treatments of the tachyon.
In section~\ref{sect.top}  we present results on the top propagator, due to 
the insertions of resummed propagators in a loop. Section~\ref{sect.bot}  deals with the bottom propagator.
In section~\ref{sect.con}  we give conclusions and outlook.
The appendices contain the relevant part of the Lagrangian
and the formulas for the one-loop integrals.

\section{One-loop self-energies at the leading order in the flavour expansion}
\label{sect.self-en}
%

In this section we shall give the expressions of the on-shell renormalized 
one-loop self-energies of the scalar particles (Higgs and would-be Goldstone bosons) at the 
 leading order in the flavour expansion. Moreover, the subtraction of tachyonic poles from the 
 Dyson resummed propagators will be presented. 
 We perform the calculation in the Landau gauge, in order to have massless unphysical scalars and we keep 
 only two mass scales, namely the top-quark, $m_t$, and the Higgs boson mass, $m_H$.

\subsection{Neutral would-be Goldstone boson} 
\label{sect.1lng}
%

In this subsection we  discuss the self-energy of the neutral would-be Goldstone boson, $\chi$, 
at the leading order in the large $N_G$-expansion. The two graphs contributing to $\Sigma_\chi(p^2)$ which are 
 enhanced by a factor $N_G$ are depicted in Fig.~\ref{fig.1} . 
 The sum of these graphs is given by (for the notation see Appendix~\ref{app.1})
\begin{figure}[h]
\begin{center}
\includegraphics[width=0.5\textwidth]{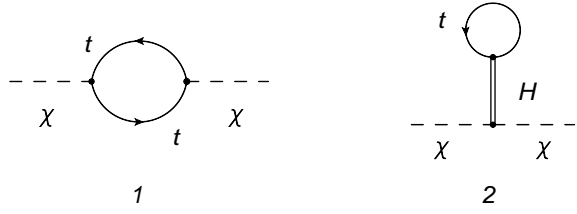}
\end{center}
\caption{SM enhanced contribution to the one-loop $\chi$ self-energy}
\label{fig.1}
\end{figure}
\begin{eqnarray}
\Sigma_\chi(p^2) = 2 i\, \sqrt{2}\,N_c\,N_G\, G_F\, m^2_t\, p^2\, B_0[p^2,m_t,m_t]\,,
\label{eq.14n}
\end{eqnarray}
where $N_c$ is the number of colours, while $N_G$ is the number of copies of the third generation.
The on-shell renormalized $\chi$ self-energy, reads 
\begin{eqnarray}
\widehat{\Sigma}_\chi(p^2) = \Sigma_\chi(p^2) -\delta m^2_\chi + \delta Z_\chi\, p^2\,,
\label{eq.15n}
\end{eqnarray}
where the mass counterterm, $\delta m^2_\chi$, and the wave function renormalization constant, 
$\delta Z_\chi$, are given by
\begin{eqnarray}
\delta m^2_\chi = \Sigma_\chi(p^2 = 0) = 0\,,~~~~
\delta Z_\chi  = -\Sigma^{'}_\chi(p^2 = 0) = 
-\alpha_t \Big[\frac{2}{D-4}+\log\Big(\frac{m_t^2}{\Lambda_B^2}\Big)+2\Big]\,.
\label{eq.16n}
\end{eqnarray}
In the above equation  we have introduced a shorthand notation $\alpha_t = \frac{\sqrt{2}}{8 \pi^2}\, N_c 
N_G\, G_F m_t^2$. We remind the reader that in the SM one has  $N_c = 3$, $N_G = 1$, and $\alpha_t = 0.0187$. 
The renormalized self-energy below the production threshold, $p^2 < 4 m_t^2$, reads
\begin{eqnarray}
\widehat{\Sigma}_\chi(p^2) = \alpha_t\,p^2 \Bigg[ 
2 \sqrt{-\Delta_\chi}\,\arctan\Bigg(\frac{1}{\sqrt{-\Delta_\chi}}\Bigg)-2\Bigg]\,, ~~~ \Delta_\chi = 1-\frac{4 m^2_t}{p^2}\,.
\label{eq.17n}
\end{eqnarray}
The expression for the on-shell renormalized $\chi$ self-energy above the production threshold is 
 given by
\begin{eqnarray}
\widehat{\Sigma}_\chi(p^2)= \alpha_t \,p^2 \Bigg\{ 
\frac{\sqrt{\Delta_\chi}}{2}\,\log\Bigg[\frac{\big(1+\sqrt{\Delta_\chi}\big)p^2-2 
m_t^2}{\big(1-\sqrt{\Delta_\chi}\big)p^2-2 m_t^2}\Bigg]-2
 -i\, \pi \,\sqrt{\Delta_\chi}\Bigg\}\,.
\label{eq.18n}
\end{eqnarray}

The behaviour of the renormalized self-energy for large momentum is:
\begin{eqnarray}
\widehat{\Sigma}_\chi(p^2) = \alpha_t\, p^2 \Big[
\log\Big(-\frac{p^2}{m^2_t} -i \epsilon\Big)-2\Big] \,,~~
 {\rm for}~~ p^2 \gg m^2_t \,. 
\label{eq.19.3}
\end{eqnarray}

\begin{figure}
\begin{center}
\includegraphics[width=0.8\textwidth]{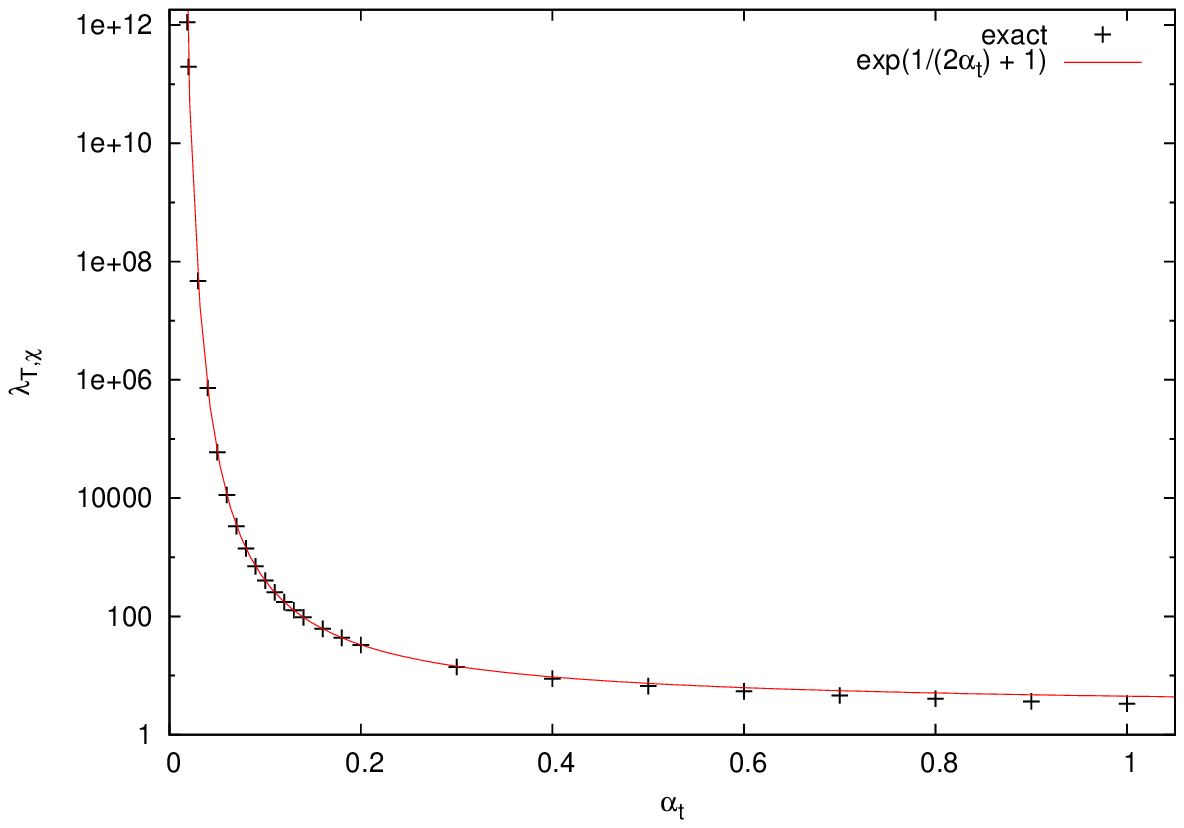}
\end{center}
\caption{Comparison between the exact result for $\lambda_{T,\chi}$ and its approximation}
\label{fig.tacn}
\end{figure}

The Dyson resummed propagator of the neutral would-be Goldstone boson $\chi$ 
at the leading order in the large $N_G$-limit is given by
\begin{eqnarray}
\widehat{D}_\chi(p^2) = \frac{i}{p^2-\widehat{\Sigma}_\chi(p^2)+i\epsilon}\,.
\label{eq.20n}
\end{eqnarray}
Besides the real pole at $p^2 = 0$ corresponding to the neutral would-be Goldstone boson, 
 the exact $\chi$ propagator in eq.(\ref{eq.20n}) contains a tachyon-pole.
 Its euclidean position, $p^2 = -\Lambda_{T,\chi}^2$, can be obtained by solving numerically the following equation
\begin{eqnarray}
\sqrt{1+\frac{4}{\lambda_{T,\chi}^2}}\, 
\log\Bigg[\frac{\lambda_{T,\chi}^2+2+\sqrt{\lambda_{T,\chi}^4+4 \lambda_{T,\chi}^2}}
{\lambda_{T,\chi}^2+2-\sqrt{\lambda_{T,\chi}^4+4 \lambda_{T,\chi}^2}}\Bigg] = \frac{2}{\alpha_t} + 4\,,
~~~ \lambda_{T,\chi}^2 = \frac{\Lambda^2_{T,\chi}}{m_t^2}\,.
\label{eq.20t}
\end{eqnarray}
A crude estimation of the position of the tachyonic pole can be given by using the approximate 
expression in eq.(\ref{eq.19.3}) instead of the full one 
\begin{eqnarray}
\Lambda_{T,\chi}^2 \simeq  m^2_t\, \exp\Big(\frac{1}{\alpha_t}+2\Big) \,.
\label{eq.21t}
\end{eqnarray}
In Fig.~\ref{fig.tacn} we show a comparison between the exact position of the tachyon (divided by the 
 top-quark mass) and the approximated expression in the above equation. 
The latter nicely reproduces the exact result for $\alpha_t < 1$, while for bigger values of the 
 coupling constant it starts overestimating it.

\begin{figure}
\begin{center}
\includegraphics[width=0.8\textwidth]{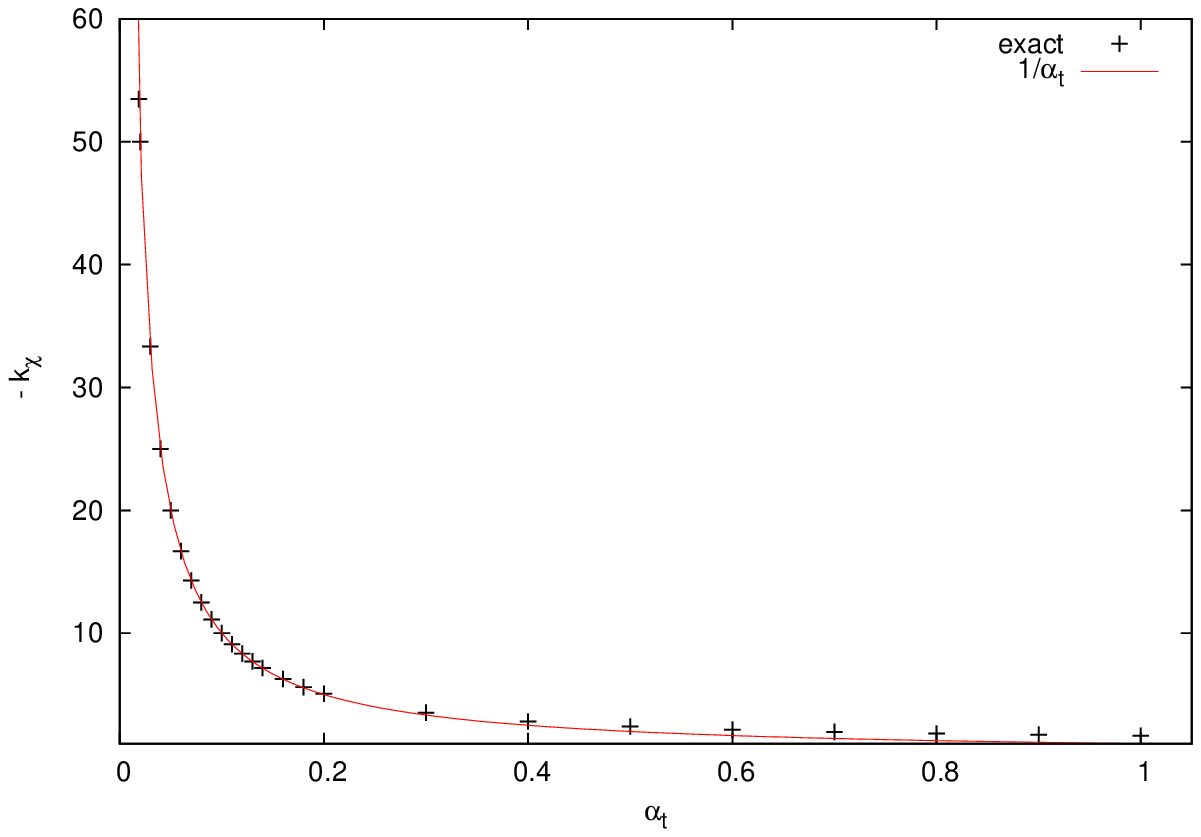}
\end{center}
\caption{Comparison between the exact result for $-\kappa_{\chi}$ and its approximation}
\label{fig.resn}
\end{figure}

The residuum at the tachyon-pole, $\kappa_\chi$, can be computed exactly in terms of $\lambda_{T,\chi}^2$
\begin{eqnarray}
\frac{1}{\kappa_{\chi}} =  -\alpha_t+
\frac{4 \alpha_t + 2}{\lambda_{T,\chi}^2+4} \,.
\label{eq.21r}
\end{eqnarray}
The opposite of the residuum, $-\kappa_{\chi}$,  is plotted against $\alpha_t$ together with 
 its approximation, $-\kappa_{\chi} = \frac{1}{\alpha_t}$, in Fig.~\ref{fig.resn}.

The spectral representation of the $\chi$ propagator (\ref{eq.20n}) is given by
\begin{eqnarray}
\widehat{D}_\chi(p^2) = \int_{-\infty}^{+\infty}\!\! ds \, \frac{i\, \rho^\chi(s)}{p^2-s+i\epsilon}\,, 
~~ {\rm where}~~ \rho^\chi (s) = \rho_T^\chi(s) + \delta(s) + \rho_+^\chi(s) \theta\big(s- 4 m^2_t\big)\,.
\label{eq.21k}
\end{eqnarray}
Notice that due to the tachyonic contribution to the spectral function,
\begin{eqnarray} 
\rho_T^\chi(s) = \kappa_\chi\, \delta\big(s+\Lambda^2_{T,\chi}\big)\,, 
\label{eq.21s}
\end{eqnarray}
the exact $\chi$ propagator (\ref{eq.20n}) does not satisfy the usual K\"allen-Lehmann spectral representation.
 The other contribution to the spectral function, which comes from the positive part of the spectrum, is given by
\begin{eqnarray} 
\rho_+^\chi(s) = \frac{\alpha_t}{s}\, \frac{\sqrt{\Delta_\chi}}{\Bigg\{1-\alpha_t 
\Bigg\{ \frac{\sqrt{\Delta_\chi}}{2}\,\log\Bigg[\frac{\big(1+\sqrt{\Delta_\chi}\big)s-2 
m_t^2}{\big(1-\sqrt{\Delta_\chi}\big)s-2 m_t^2}\Bigg]-2 \Bigg\} \Bigg\}^{\!\!2} + \pi^2\, \alpha_t^2\, 
\Delta_\chi}\,. 
\label{eq.21p}
\end{eqnarray}
The integral over the spectrum of $\rho_+^\chi$ is convergent, since in the high-energy limit one has
\begin{eqnarray} 
\rho_+^\chi(s) \simeq \frac{\alpha_t}{s}\, \frac{1}{\Big[1-\alpha_t 
\log\Big(\frac{s}{m_t^2}\Big)+2 \alpha_t \Big]^2 + \pi^2 \alpha_t^2}\,.
\label{eq.21pbis}
\end{eqnarray} 
Notice that this is not the case in perturbation theory.  
Indeed, if one expands eq.(\ref{eq.21p}) in powers of  $\alpha_t$, the resulting spectral function at the 
 leading perturbative order, 
\begin{eqnarray} 
\rho_+^\chi(s) = \frac{\alpha_t}{s}\, \sqrt{1-\frac{4m^2_t}{s}} + O\big(\alpha_t\big)^2\,,
\label{eq.21pter}
\end{eqnarray} 
is clearly not integrable over the positive part of the spectrum. Thus, the resummation provides a 
cut-off to the theory.

By using the residue theorem, one can prove that the integral over the whole spectrum 
 of the spectral function, $\rho^\chi$, vanishes. 

\begin{eqnarray}
\int _{-\infty}^{+\infty} \!\! ds\,\, \rho^\chi(s) = \kappa_\chi +1 +
\int _{4 m^2_t}^{+\infty} \!\! ds\,\, \rho_+^\chi(s) = 0\,. 
\label{eq.21i}
\end{eqnarray}
The above result can also be checked with a careful numerical integration.

Clearly, the removal of the tachyonic pole is necessary in order to find an expression for the 
 resummed $\chi$ propagator that respects causality and satisfies the K\"allen-Lehmann 
 representation. On the other hand, the contribution of the tachyon-pole is crucial in order to 
 ensure the normalization of the spectral function in eq.(\ref{eq.21i}).
We propose to minimally subtract the tachyonic pole 

\begin{eqnarray}
\widehat{D}^{{\rm MS}}_\chi(p^2) = \frac{i}{p^2-\widehat{\Sigma}_\chi(p^2)+i\epsilon}- 
\frac{i\, \kappa_\chi}{p^2 + \Lambda^2_{\chi,T}}\,.
\label{eq.21sub}
\end{eqnarray}
Furthermore, one can impose the condition that the integral over the physical, subtracted, spectral density 
 be equal to one. This amounts to rescaling the subtracted propagator by a factor $-\frac{1}{\kappa_\chi}$. 
 We call this prescription the Akhoury scheme in the following \cite{akhoury}.

Another possibility is to perform a non-minimal subtraction of the tachyon (for a similar strategy in the context 
of QCD see Refs.\cite{bMS}).  One can, for instance, subtract the tachyonic pole and add its residuum to the pion pole at $p^2 = 0$.
This prescription will be called beyond-the-minimal-subtraction (bMS) scheme.
\begin{eqnarray}
\widehat{D}^{{\rm bMS}}_\chi(p^2) = \frac{i}{p^2-\widehat{\Sigma}_\chi(p^2)+i\epsilon} +
\frac{i\, \kappa_\chi\, \Lambda^2_{\chi,T}}{p^2\big(p^2 + \Lambda^2_{\chi,T}\big)}\,.
\label{eq.21com.3}
\end{eqnarray}
This solution has the  property of removing the tachyon without modifying the normalization 
of the spectral function. Indeed, one finds:
\begin{eqnarray}
\widehat{D}^{{\rm bMS}}_\chi(p^2) = 
\int_{0}^{+\infty}\!\! ds \, \frac{i\, \rho_{{\rm bMS}}^\chi(s)}{p^2-s+i\epsilon}\,,
~~ {\rm where}~~ \rho_{{\rm bMS}}^\chi (s) = \big(1+\kappa_\chi\big) \delta(s) + 
\rho_+^\chi(s) \theta\big(s- 4 m^2_t\big)\,.
\label{eq.21com.4}
\end{eqnarray}
We remark that there are other non-minimal ways of removing the tachyon. Another choice could be to impose the validity of the 
tree-level relation, i.e. integral of the spectral function equal to one. In the latter case the subtraction term is such that the 
resulting spectral function is given by:
\begin{eqnarray}
\rho^\chi (s) = \big(2+\kappa_\chi\big) \delta(s) +
\rho_+^\chi(s) \theta\big(s- 4 m^2_t\big)\,. 
\label{eq.news}
\end{eqnarray}
However, due to the lack of a 
subsidiary principle, like asymptotic freedom in QCD, we adopt, for the sake of simplicity, the minimal 
subtraction prescription in our computation of renormalon contribution to the top- and bottom-quark propagators.

\subsection{Charged would-be Goldstone boson}
\label{sect.1lcg}

In this subsection we shall discuss the self-energy of the charged would-be Goldstone boson, $\phi$,
at the leading order in the large $N_G$-expansion. The two graphs contributing to $\Sigma_\phi(p^2)$ which are
 enhanced by a factor $N_G$ are depicted in Fig.~\ref{fig.2}. 
The sum of these graphs is given by (for the notation see Appendix~\ref{app.1})
\begin{figure}[h]
\begin{center}
\includegraphics[width=0.5\textwidth]{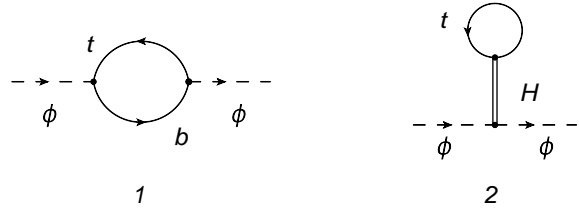}
\end{center}
\caption{SM enhanced contribution to the one-loop $\phi$ self-energy}
\label{fig.2}
\end{figure}
\begin{eqnarray}
\Sigma_\phi(p^2) = 2 i\, \sqrt{2}\,N_c\,N_G\, G_F\, m^2_t\Big[\big(p^2-m^2_t\big) B_0[p^2,m_t,0]
+  A_0[m_t]\Big] \,.
\label{eq.21n}
\end{eqnarray}
The on-shell renormalized $\phi$ self-energy reads 
\begin{eqnarray}
\widehat{\Sigma}_\phi(p^2) = \Sigma_\phi(p^2) -\delta m^2_\phi +\delta Z_\phi \,p^2\,,
\label{eq.22n}
\end{eqnarray}
where the mass counterterm, $\delta m^2_\phi$, and the wave function renormalization constant,
$\delta Z_\phi$, are given by
\begin{eqnarray}
\delta m^2_\phi = \Sigma_\phi(p^2 = 0) = 0\,,~~~~
\delta Z_\phi  = -\Sigma^{'}_\phi(p^2 = 0) = - \alpha_t 
\Big[\frac{2}{D-4}+\log\Big(\frac{m_t^2}{\Lambda_B^2}\Big)+\frac{3}{2}\Big]\,.
\label{eq.23n}
\end{eqnarray}
The renormalized self-energy below the production threshold, $p^2 < m_t^2$, reads
\begin{eqnarray}
\widehat{\Sigma}_\phi(p^2) = \alpha_t\, p^2\,  \Bigg[
\Delta^2_\phi\, \log\Big(\frac{-\Delta_\phi}{1-\Delta_\phi}\Big)- \Delta_\phi
-\frac{1}{2} \Bigg]\,,~~~ \Delta_\phi = 1-\frac{m_t^2}{p^2}\,.
\label{eq.24n}
\end{eqnarray}
The expression for the on-shell renormalized $\phi$ self-energy above the production threshold is
 given by
\begin{eqnarray}
\widehat{\Sigma}_\phi(p^2) = \alpha_t\,p^2\, \Bigg[
\Delta^2_\phi\, \log\Big(\frac{\Delta_\phi}{1-\Delta_\phi}\Big)- \Delta_\phi-\frac{1}{2} - i \pi \, \Delta^2_\phi
\Bigg]\,.
\label{eq.25n}
\end{eqnarray}

The behaviour of the renormalized self-energy for large momentum is:
\begin{eqnarray}
\widehat{\Sigma}_\phi(p^2) = \alpha_t\, p^2 \Big[
\log\Big(-\frac{p^2}{m^2_t} -i \epsilon\Big)-\frac{3}{2} \Big] \,,~~
 {\rm for}~~ p^2 \gg m^2_t \,.
\label{eq.26.3}
\end{eqnarray}

The Dyson resummed propagator of the charged would-be Goldstone boson $\phi$
at the leading order in the large $N_G$-limit is given by
\begin{eqnarray}
\widehat{D}_\phi(p^2) = \frac{i}{p^2-\widehat{\Sigma}_\phi(p^2)+i\epsilon}\,.
\label{eq.27n}
\end{eqnarray}

\begin{figure}
\begin{center}
\includegraphics[width=0.8\textwidth]{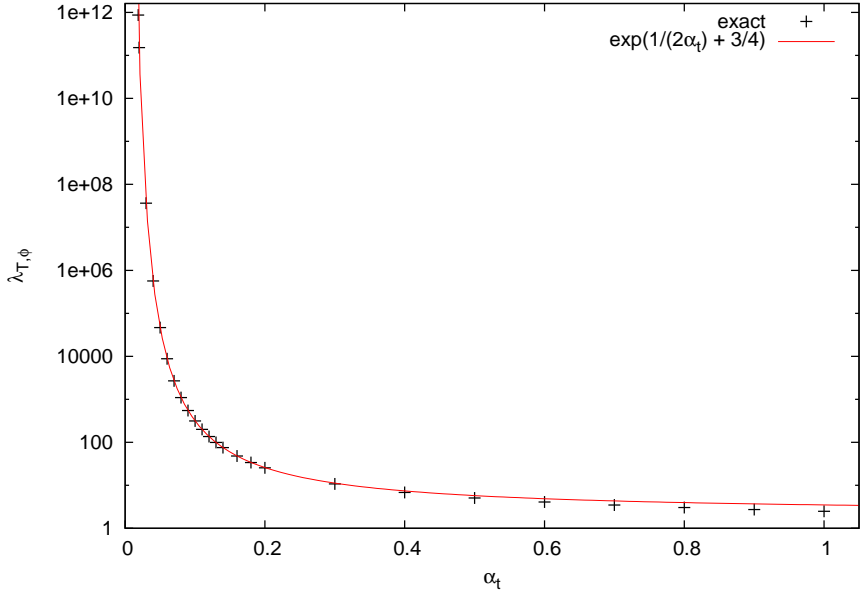}
\end{center}
\caption{Comparison between the exact result for $\lambda_{T,\phi}$ and its approximation}
\label{fig.tacc}
\end{figure}

Besides the real pole at $p^2 = 0$ corresponding to the charged would-be Goldstone boson, 
 the exact $\phi$ propagator in eq.(\ref{eq.27n}) contains a tachyon-pole.
 Its euclidean position, $p^2 = -\Lambda_{T,\phi}^2$, can be obtained by solving numerically the following equation
\begin{eqnarray}
\Big(1+\frac{1}{\lambda_{T,\phi}^2}\Big)^2\,
\Big[\log\big(\lambda^2_{T,\phi} + 1\big)- \frac{\lambda_{T,\phi}^2}{\lambda_{T,\phi}^2+1}\Big] = 
\frac{1}{\alpha_t} + \frac{1}{2}\,,
~~~ \lambda_{T,\phi}^2 = \frac{\Lambda^2_{T,\phi}}{m_t^2}\,.
\label{eq.27t}
\end{eqnarray}
A crude estimation of the position of the tachyonic pole can be given by using the approximate expression in eq.(\ref{eq.26.3})
 instead of the full one
\begin{eqnarray}
\Lambda_{T,\phi}^2 \simeq  m^2_t\, \exp\Big(\frac{1}{\alpha_t}+\frac{3}{2}\Big) \,.
\label{eq.28t}
\end{eqnarray}
In Fig.~\ref{fig.tacc} we show a comparison between the exact position of the tachyon (divided by the
 top-quark mass) and the approximated expression in the above equation. 
The latter nicely reproduces the exact result for $\alpha_t < 1$, while for bigger values of the coupling constant it 
 starts overestimating it.

\begin{figure}
\begin{center}
\includegraphics[width=0.8\textwidth]{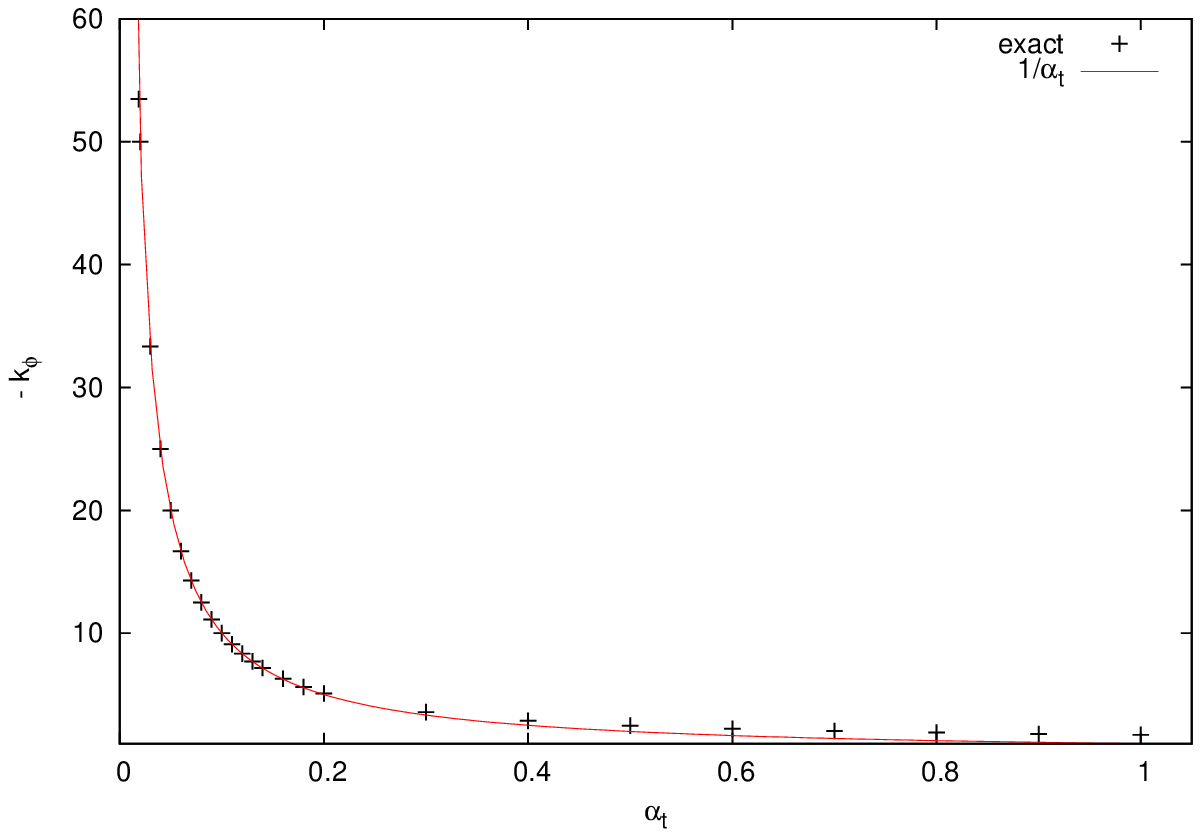}
\end{center}
\caption{Comparison between the exact result for $-\kappa_{\phi}$ and its approximation}
\label{fig.resc}
\end{figure}

The residuum at the tachyon-pole, $\kappa_\phi$, can be computed exactly in terms of $\lambda_{T,\phi}^2$
\begin{eqnarray}
\frac{1}{\kappa_\phi} = -\alpha_t + \frac{\alpha_t+2}{\lambda^2_{T,\phi}+1}\,.
\label{eq.28r}
\end{eqnarray}
The opposite of the residuum, $-\kappa_{\phi}$,  is plotted against $\alpha_t$ together with
 its approximation, $-\kappa_{\phi} = \frac{1}{\alpha_t}$ in Fig.~\ref{fig.resc}.

The  spectral representation of the $\phi$ propagator (\ref{eq.27n}) is given by
\begin{eqnarray}
\widehat{D}_\phi(p^2) = \int_{-\infty}^{+\infty}\!\! ds \, \frac{i\, \rho^\phi(s)}{p^2-s+i\epsilon}\,,
~~ {\rm where}~~ \rho^\phi (s) = \rho_T^\phi(s) + \delta(s) + \rho_+^\phi(s) \theta\big(s-  m^2_t\big)\,.
\label{eq.28k}
\end{eqnarray}
Notice that due to the tachyonic contribution to the spectral function,
\begin{eqnarray}
\rho_T^\phi(s) = \kappa_\phi\, \delta\big(s+\Lambda^2_{T,\phi}\big)\,,
\label{eq.28s}
\end{eqnarray}
the exact $\phi$ propagator (\ref{eq.27n}) does not satisfy the usual K\"allen-Lehmann spectral representation.
 The other contribution to the spectral function, which comes from the positive part of the spectrum, is given by
\begin{eqnarray}
\rho_+^\phi(s) = \frac{\alpha_t}{s}\, \frac{\Delta^2_\phi}{\Big\{1-\alpha_t
\Big[\Delta^2_\phi\,\log\Big(\frac{\Delta_\phi}{1-\Delta_\phi}\Big)-\Delta_\phi-\frac{1}{2}\Big] \Big\}^2 + \pi^2\, 
\alpha_t^2\, \Delta^4_\phi}\,.
\label{eq.28p}
\end{eqnarray}
The integral over the spectrum of $\rho_+^\phi$ is convergent, since in the high energy limit one has
\begin{eqnarray}
\rho_+^\phi(s) \simeq \frac{\alpha_t}{s}\, \frac{1}{\Big[1-\alpha_t
\log\Big(\frac{s}{m_t^2}\Big)+ \frac{3}{2} \alpha_t \Big]^2 + \pi^2 \alpha_t^2}\,.
\label{eq.28pbis}
\end{eqnarray}
Notice that, again, this is not the case in perturbation theory.  
By expanding the spectral function in eq.(\ref{eq.28p}) in powers of $\alpha_t$, one gets a function, 
\begin{eqnarray}
\rho_+^\phi(s) = \frac{\alpha_t}{s}\, \Big(1-\frac{m^2_t}{s}\Big)^2 + O\big(\alpha_t^2\big)\,,
\label{eq.28pter}
\end{eqnarray}
which is not integrable over the positive part of the spectrum.

By using the residue theorem, one can prove that the integral over the whole spectrum 
of $\rho^\phi$ vanishes
\begin{eqnarray}
\int _{-\infty}^{+\infty} \!\! ds\,\, \rho^\phi(s) = \kappa_\phi +1 +
\int _{m^2_t}^{+\infty} \!\! ds\,\, \rho_+^\phi(s) = 0\,.
\label{eq.28i}
\end{eqnarray}
The above result has been confirmed by a careful numerical integration.

As in the neutral case we minimally subtract the tachyonic pole 

\begin{eqnarray}
\widehat{D}^{{\rm MS}}_\phi(p^2) = \frac{i}{p^2-\widehat{\Sigma}_\phi(p^2)+i\epsilon}-
 \frac{i\, \kappa_\phi}{p^2 + \Lambda^2_{\phi,T}}\,.
\label{eq.28sub}
\end{eqnarray}

Normalizing the spectral density  amounts to rescaling the subtracted propagator by a factor $-\frac{1}{\kappa_\phi}$. 

In the case of the bMS scheme
the tachyon-subtracted propagator is given by:
\begin{eqnarray}
\widehat{D}^{{\rm bMS}}_\phi(p^2) = \frac{i}{p^2-\widehat{\Sigma}_\phi(p^2)+i\epsilon} +
\frac{i\, \kappa_\phi\, \Lambda^2_{\phi,T}}{p^2\big(p^2 + \Lambda^2_{\phi,T}\big)}\,.
\label{eq.28com.3}
\end{eqnarray}
Hereby one removes the tachyon, but keeps the spectral density normalized. One finds:
\begin{eqnarray}
\widehat{D}^{{\rm bMS}}_\phi(p^2) = 
\int_{0}^{+\infty}\!\! ds \, \frac{i\, \rho_{{\rm bMS}}^\phi(s)}{p^2-s+i\epsilon}\,,
~~ {\rm where}~~ \rho_{{\rm bMS}}^\phi (s) = \big(1+\kappa_\phi\big) \delta(s) + 
\rho_+^\phi(s) \theta\big(s- m^2_t\big)\,.
\label{eq.28com.4}
\end{eqnarray}
%

\subsection{Neutral Higgs boson}
\label{sect.1lh}

In this subsection we shall discuss the self-energy of the neutral Higgs boson
at the leading order in the large $N_G$-expansion. We consider a finite, but not completely arbitrary Higgs mass, 
namely $m_H < 2 m_t$. In this way the Higgs boson cannot decay in $t\, \bar{t}$ and thus, it is 
 stable at the leading order in the large $N_G$-limit. 
The two graphs contributing to $\Sigma_H(p^2)$ which are
 enhanced by a factor $N_G$ are depicted in Fig.~\ref{fig.3}. 
 The sum of these graphs is given by (for the notation see Appendix~\ref{app.1})
\begin{figure}[h]
\begin{center}
\includegraphics[width=0.5\textwidth]{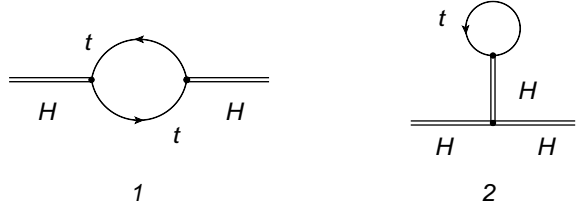}
\end{center}
\caption{SM enhanced contribution to the one-loop Higgs self-energy}
\label{fig.3}
\end{figure}
\begin{eqnarray}
\Sigma_H(p^2) = 2 i\, \sqrt{2}\,N_c\,N_G\, G_F\, m^2_t\Big[\big(p^2-4m^2_t\big)\, B_0[p^2,m_t,m_t]
+4 A_0[m_t]\Big] \,,
\label{eq.7n}
\end{eqnarray}
The on-shell renormalized Higgs self-energy reads
\begin{eqnarray}
\widehat{\Sigma}_H(p^2) = \Sigma_H(p^2) -\delta m^2_H +\delta Z_H \big(p^2-m^2_H\big)\,,
\label{eq.8n}
\end{eqnarray}
where the mass counterterm, $\delta m^2_H$, and the wave function renormalization constant, 
$\delta Z_H$, are given by
\begin{eqnarray}
&&\!\!\!\!\!\!\!\!\!\!\!\!
\delta m^2_H = \Sigma_H(p^2= m^2_H) = \alpha_t\, \Big\{ m^2_H \Big[
\frac{2}{D-4} + \log\Big(\frac{m_t^2}{\Lambda_B^2}\Big) + 2 \sqrt{-\Delta_H} 
\arctan\Big(\frac{1}{\sqrt{-\Delta_H}}\Big)\Big] \nonumber\\
&&~~~~~~~~~~~~~~~~~~~~~~~~~
+ 4 m_t^2 \Big[1 - 2 \sqrt{-\Delta_H} \arctan\Big(\frac{1}{\sqrt{-\Delta_H}}\Big)\Big]\Big\} 
\,,~~ \Delta_H = 1 - \frac{4 m_t^2}{m^2_H}\nonumber\\
&&\!\!\!\!\!\!\!\!\!\!\!\!
\delta Z_H  = - \Sigma^{'}_H(p^2 = m_H^2) = -\alpha_t 
\Big[\frac{2}{D-4} + \log\Big(\frac{m_t^2}{\Lambda_B^2}\Big) + \Delta_H + 
2\Big(1+\frac{2 m_t^2}{m_H^2}\Big) \sqrt{-\Delta_H}\times \nonumber\\
&&~~~~~~~~~~~~~~~~~~~~~~~~~~~~~~~~~~
\arctan\Big(\frac{1}{\sqrt{-\Delta_H}}\Big)\Big]\,.
\label{eq.9n}
\end{eqnarray}
In the limit of vanishing Higgs mass, $m_H = 0$, one finds
\begin{eqnarray}
\delta m_H^2 = -4 \alpha_t\, m^2_t\,,~~~~ \delta Z_H = -\alpha_t\Big[\frac{2}{D-4} + 
\log\Big(\frac{m_t^2}{\Lambda_B^2}\Big) + \frac{8}{3}\Big]\,.
\label{eq.9n.1}
\end{eqnarray}
It is interesting to notice that there is a finite Higgs mass renormalization even if one neglets $m_H$. 
 This effect comes from the v.e.v. of the Higgs field.


The renormalized Higgs self-energy below the production threshold, i.e. for $p^2 < 4 m^2_t$, reads 
\begin{eqnarray}
\widehat{\Sigma}_H(p^2) &\!\!\!= &\!\!\! \alpha_t\, p^2 \Bigg\{2 \Delta_\chi \Bigg[ \sqrt{-\Delta_\chi}
\,\arctan\Bigg(\frac{1}{\sqrt{-\Delta_\chi}}\Bigg)-\sqrt{-\Delta_H}\,
\arctan\Bigg(\frac{1}{\sqrt{-\Delta_H}}\Bigg)\Bigg] \nonumber\\
&&~~~~~~
-\Big(1-\frac{m_H^2}{p^2}\Big)
\Bigg[\Delta_H + \frac{4 m_t^2}{m_H^2}\, 
\sqrt{-\Delta_H}\,\arctan\Bigg(\frac{1}{\sqrt{-\Delta_H}}\Bigg)\Bigg]\Bigg\}\,.
\label{eq.10n}
\end{eqnarray}
The above expression simplifies a lot in the limit of vanishing Higgs mass
\begin{eqnarray}
\widehat{\Sigma}_H(p^2) = \alpha_t\, p^2 \Bigg[2 \Delta_\chi\, \sqrt{-\Delta_\chi}
\,\arctan\Bigg(\frac{1}{\sqrt{-\Delta_\chi}}\Bigg)-2 \Delta_\chi -\frac{2}{3}\Bigg]\,.
\label{eq.10n.1}
\end{eqnarray}
The expression for the on-shell renormalized Higgs self-energy above the production threshold is  given by
\begin{eqnarray}
\widehat{\Sigma}_H(p^2) &\!\!\!= &\!\!\! \alpha_t\, p^2 \Bigg\{\Delta_\chi \Bigg\{ 
\frac{\sqrt{\Delta_\chi}}{2}\,\log\Bigg[\frac{\big(1+\sqrt{\Delta_\chi}\big)p^2-2
m_t^2}{\big(1-\sqrt{\Delta_\chi}\big)p^2-2 m_t^2}\Bigg]
-2 \sqrt{-\Delta_H}\,\arctan\Bigg(\frac{1}{\sqrt{-\Delta_H}}\Bigg)  \nonumber\\
&&~~~\!
 - i \pi\, \sqrt{\Delta_\chi} \Bigg\} -\Big(1-\frac{m_H^2}{p^2}\Big)
\Bigg[\Delta_H + \frac{4 m_t^2}{m_H^2}\,
\sqrt{-\Delta_H}\,\arctan\Bigg(\frac{1}{\sqrt{-\Delta_H}}\Bigg)\Bigg]\Bigg\}\,.
\label{eq.11n}
\end{eqnarray}
Also in this case, if we neglect the Higgs mass we obtain a simplified expression
\begin{eqnarray}
\widehat{\Sigma}_H(p^2)= \alpha_t\, p^2 \Bigg\{\frac{\Delta_\chi}{2}\, 
\sqrt{\Delta_\chi}\,\log\Bigg[\frac{\big(1+\sqrt{\Delta_\chi}\big)p^2-2 
m_t^2}{\big(1-\sqrt{\Delta_\chi}\big)p^2-2 m_t^2}\Bigg]-2 \Delta_\chi -\frac{2}{3}
- i\, \pi \, \Delta_\chi\, \sqrt{\Delta_\chi}\Bigg\}\,.
\label{eq.11n.1}
\end{eqnarray}

The behaviour of the renormalized self-energy for large momentum is: 
\begin{eqnarray}
\label{eq.12.1}
\widehat{\Sigma}_H(p^2) = \alpha_t\,p^2\Big[
\log\Big(-\frac{p^2}{m^2_t} -i \epsilon\Big)-\frac{8}{3}\Big] \,,~~
 {\rm for}~~ p^2 \gg m^2_t\,, m^2_H \,. 
\label{eq.12.3}
\end{eqnarray}

The Dyson resummed Higgs propagator at the leading order in the large $N_G$-limit is
 given by
\begin{eqnarray}
\widehat{D}_H(p^2) = \frac{i}{p^2-m^2_H-\widehat{\Sigma}_H(p^2)+i\epsilon}\,.
\label{eq.13n}
\end{eqnarray}
Besides the real pole at $p^2 = m_H^2$ corresponding to the Higgs particle, 
 the exact Higgs propagator in eq.(\ref{eq.13n}) contains a tachyon-pole.
Its euclidean position, $p^2 = -\Lambda^2_{T,H}$, can be obtained by solving numerically 
 the following equation 
%
\begin{eqnarray}
&&\!\!\!\!\!\!\!\!\!\!\!\!\!\!\!\!\!
\Bigg(1+\frac{4}{\lambda_{T,H}^2}\Bigg)\Bigg\{\sqrt{1+\frac{4}{\lambda_{T,H}^2}}\,
\log\Bigg[\frac{\lambda_{T,H}^2+2+\sqrt{\lambda_{T,H}^4+4 \lambda_{T,H}^2}}
{\lambda_{T,H}^2+2-\sqrt{\lambda_{T,H}^4+4 \lambda_{T,H}^2}}\Bigg]-4 \sqrt{-\Delta_H}\,
\arctan\Bigg(\frac{1}{\sqrt{-\Delta_H}}\Bigg)\Bigg\}\nonumber\\
&&\!\!\!\!\!\!\!\!\!\!\!\!\!\!
= \Bigg(2+\frac{m^2_H}{m_t^2}\,\frac{2}{\lambda_{T,H}^2}\Bigg)\Bigg[ \frac{1}{\alpha_t} +
\Delta_H+ \frac{4 m^2_t}{m_H^2}\, \sqrt{-\Delta_H}\,
\arctan\Bigg(\frac{1}{\sqrt{-\Delta_H}}\Bigg)\Bigg]\,,~~ \lambda_{T,H}^2 = \frac{\Lambda^2_{T,H}}{m_t^2}\,.
\label{eq.13t}
\end{eqnarray}
In the zero Higgs mass limit the above equation simplifies and reads
\begin{eqnarray}
\Bigg(1+\frac{4}{\lambda_{T,H}^2}\Bigg)\Bigg\{\sqrt{1+\frac{4}{\lambda_{T,H}^2}}\,
\log\Bigg[\frac{\lambda_{T,H}^2+2+\sqrt{\lambda_{T,H}^4+4 \lambda_{T,H}^2}}
{\lambda_{T,H}^2+2-\sqrt{\lambda_{T,H}^4+4 \lambda_{T,H}^2}}\Bigg]-4\Bigg\} = \frac{2}{\alpha_t} + 
\frac{4}{3}\,.
\label{eq.13tbis}
\end{eqnarray}
The impact of a finite Higgs mass on the position of the tachyon can be quite sizable. 
Indeed, it turns out that with a finite Higgs mass, $m_H = 125\,$ GeV and $m_t = 172\,$ GeV as a reference top 
mass, $\lambda_{T,H}$ is about $6\, \%\, -\, 8 \,\%$ smaller than the same quantity with zero Higgs mass.  

\begin{figure}
\begin{center}
\includegraphics[width=0.8\textwidth]{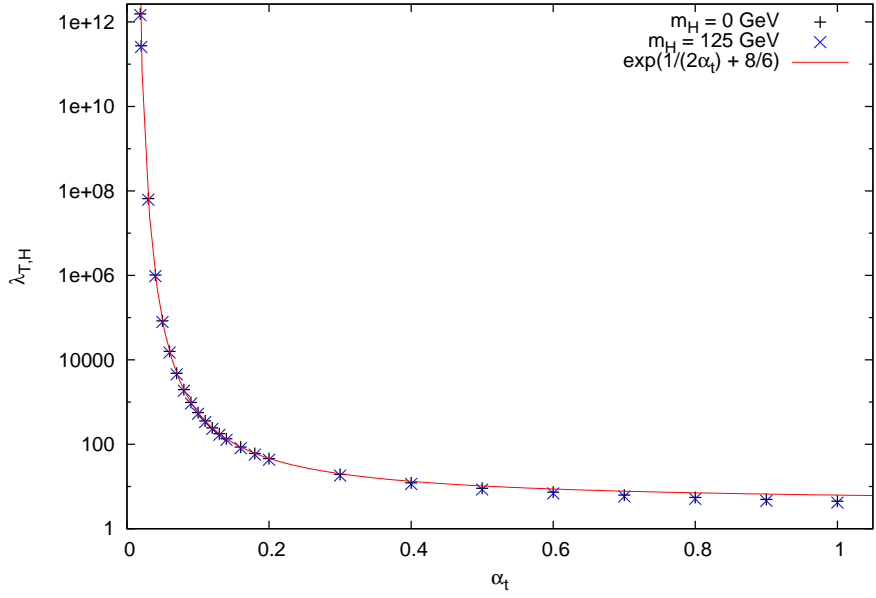}
\end{center}
\caption{Comparison between the exact result for $\lambda_{T,H}$ and its approximation}
\label{fig.tach}
\end{figure}

A crude estimation of the position of the tachyonic pole can be given by using the approximate expression in eq.(\ref{eq.12.3})
 instead of the full one
\begin{eqnarray}
\Lambda_{T,H}^2 = m^2_t\, \exp\Big(\frac{1}{\alpha_t}+\frac{8}{3}\Big) \,.
\label{eq.14t}
\end{eqnarray}
In Fig.~\ref{fig.tach} we show a comparison between the exact position of the tachyon (divided by the
 top-quark mass) both for a massless Higgs boson and for $m_H = 125\,$ GeV 
and the approximated expression in the above equation. 
The latter nicely reproduces the exact result for $\alpha_t < 1$, while for bigger values of the coupling constant it 
 starts overestimating it.
%

\begin{figure}
\begin{center}
\includegraphics[width=0.8\textwidth]{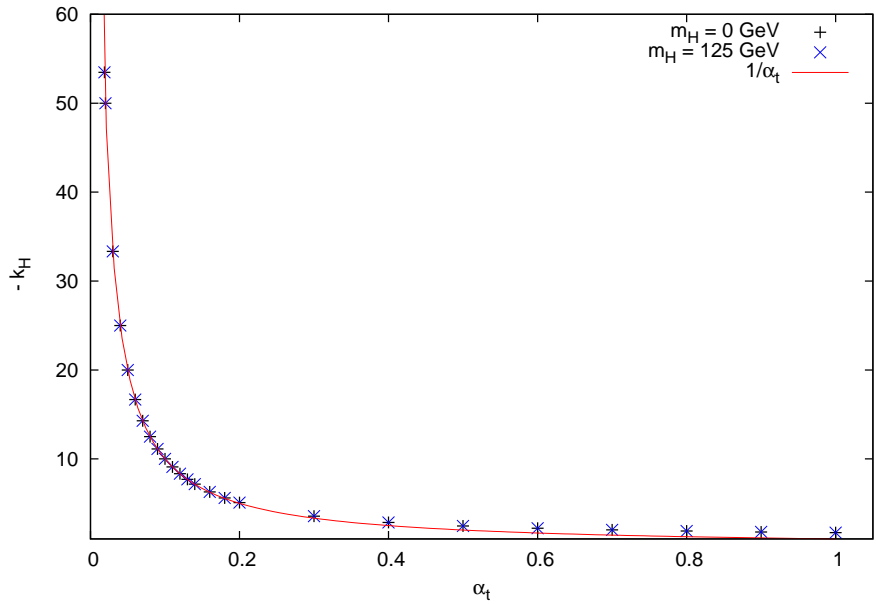}
\end{center}
\caption{Comparison between the exact result for $-\kappa_{H}$ and its approximation}
\label{fig.resh}
\end{figure}

The residuum at the tachyon-pole, $\kappa_H$, can be computed exactly in terms of $\lambda_{T,H}^2$ and of the 
 ratio $m_H^2/m_t^2$.
\begin{eqnarray}
\frac{1}{\kappa_H} \!\!\! & = \!\!\! &  1-\Bigg(1+\frac{m^2_H}{m_t^2}\,\frac{1}{\lambda_{T,H}^2}\Bigg)\Bigg\{
1+\alpha_t-\frac{6 \alpha_t}{\lambda_{T,H}^2+4}\Bigg[ \frac{1}{\alpha_t} +
\Delta_H 
+ \frac{4 m^2_t}{m_H^2}\, \sqrt{-\Delta_H} \times 
\nonumber\\
&&~~~~~~~~~~~~~~~~~~~~~~~~~~~~~
\arctan\Bigg(\frac{1}{\sqrt{-\Delta_H}}\Bigg)\Bigg]\Bigg\}\,.
\label{eq.13r}
\end{eqnarray}
In the limit of vanishing Higgs mass the above equation simplifies and reads
\begin{eqnarray}
\frac{1}{\kappa_H} = -\alpha_t + \frac{4 \alpha_t + 6}{\lambda^2_{T,H}+4}\,.
\label{eq.13rbis}
\end{eqnarray}
In the above equation $\lambda^2_{T,H}$ is the solution of eq.(\ref{eq.13tbis}) and not of the complete equation.
 We found that the impact of a finite Higgs mass on the residuum at the tachyonic pole is completely negligible.
The opposite of the residuum, $-\kappa_{H}$, is plotted against $\alpha_t$ together with
 its approximation, $-\kappa_{H} = \frac{1}{\alpha_t}$ in Fig.~\ref{fig.resh}.

The  spectral representation of the Higgs propagator (\ref{eq.13n}) is given by
\begin{eqnarray}
\widehat{D}_H(p^2) = \int_{-\infty}^{+\infty}\!\! ds \, \frac{i\, \rho^H(s)}{p^2-s+i\epsilon}\,,
~~ {\rm where}~~ \rho^H (s) = \rho_T^H(s) + \delta(s-m^2_H) + \rho_+^H(s) \theta\big(s- 4 m^2_t\big)\,.
\label{eq.13k}
\end{eqnarray}
Notice that due to the tachyonic contribution to the spectral function,
\begin{eqnarray}
\rho_T^H(s) = \kappa_H\, \delta\big(s+\Lambda^2_{T,H}\big)\,,
\label{eq.13s}
\end{eqnarray}
the exact Higgs propagator (\ref{eq.13n}) does not satisfy the usual K\"allen-Lehmann spectral representation.
 The other contribution to the spectral function, which comes from the positive part of the spectrum, is given by
\begin{eqnarray}
\rho_+^H(s) = \frac{\alpha_t}{s}\, \frac{\Delta_\chi\, \sqrt{\Delta_\chi}}{\Bigg\{1-\alpha_t
\Bigg\{ \frac{\Delta_\chi}{2}\,\sqrt{\Delta_\chi} \log\Bigg[\frac{\big(1+\sqrt{\Delta_\chi}\big)s-2
m_t^2}{\big(1-\sqrt{\Delta_\chi}\big)s-2 m_t^2}\Bigg]-2 \Delta_\chi-\frac{2}{3} \Bigg\} \Bigg\}^{\!\!2} + \pi^2\, 
\alpha_t^2\,\Delta^3_\chi}\,.
\label{eq.13p}
\end{eqnarray}
The integral over the spectrum of $\rho_+^H$ is convergent, since in the high energy limit one has
\begin{eqnarray}
\rho_+^H(s) \simeq \frac{\alpha_t}{s}\, \frac{1}{\Big[1-\alpha_t
\log\Big(\frac{s}{m_t^2}\Big)+ \frac{8}{3} \alpha_t \Big]^2 + \pi^2 \alpha_t^2}\,.
\label{eq.13pbis}
\end{eqnarray}
In the above equations we have reported the positive part of the spectral function
 for a massless Higgs boson. 
The complete expression of $\rho_+^H$ for a generic Higgs mass is rather  
cumbersome and can be easily obtained from eq.(\ref{eq.11n}).  

Notice that also for the Higgs boson the integral over the spectrum is divergent order by order 
in perturbation theory. Indeed, by expanding $\rho_+^H$ in powers of $\alpha_t$, one finds, at 
the leading order, a function,
\begin{eqnarray}
\rho_+^H(s) = \frac{\alpha_t}{s}\, \Big(1-\frac{4 m^2_t}{s}\Big)\, \sqrt{1-\frac{4 m^2_t}{s}} + O\big(\alpha_t^2\big)\,,
\label{eq.13pter}
\end{eqnarray}
which is not integrable over the positive part of the spectrum.

By using the residue theorem, one can prove that the integral over the whole spectrum of 
 $\rho^H$ vanishes (both for a massless and a massive Higgs boson)
\begin{eqnarray}
\int _{-\infty}^{+\infty} \!\! ds\,\, \rho^H(s) = \kappa_H +1 +
\int _{4 m^2_t}^{+\infty} \!\! ds\,\, \rho_+^H(s) = 0\,.
\label{eq.13i}
\end{eqnarray}
The above result has been confirmed by a careful numerical integration.

We propose to minimally subtract the tachyonic pole 
\begin{eqnarray}
\widehat{D}^{{\rm MS}}_H(p^2) = \frac{i}{p^2-m_H^2-\widehat{\Sigma}_H(p^2)+i\epsilon}-
\frac{i\, \kappa_H}{p^2 + \Lambda^2_{H,T}} \,.
\label{eq.13sub}
\end{eqnarray}
Normalizing the spectral density amounts to rescaling the subtracted propagator by a factor $-\frac{1}{\kappa_H}$. 

According to the bMS scheme the tachyon-subtracted propagator is given by:
\begin{eqnarray}
\widehat{D}^{{\rm bMS}}_H(p^2) = \frac{i}{p^2-m_H^2-\widehat{\Sigma}_H(p^2)+i\epsilon} +
\frac{i\, \kappa_H\, \big(\Lambda^2_{H,T}+m_H^2\big)}{\big(p^2-m_H^2\big)\big(p^2 + \Lambda^2_{H,T}\big)}\,.
\label{eq.13com.3}
\end{eqnarray}
Also here one removes the tachyon without modifying the normalization
of the spectral function. One finds:
\begin{eqnarray}
\!
\widehat{D}^{{\rm bMS}}_H(p^2) = \!\!
\int_{0}^{+\infty}\!\!\!\!\!\!\! ds \, \frac{i\, \rho_{{\rm bMS}}^H(s)}{p^2-s+i\epsilon}\,,
\,\, {\rm where}\,\, \rho_{{\rm bMS}}^H (s) = \big(1+\kappa_H\big) \delta(s-m_H^2) +
\rho_+^H(s) \theta\big(s- 4 m^2_t\big)\,.
\label{eq.13com.4}
\end{eqnarray}
By using the non-minimal subtraction term in eq.(\ref{eq.13com.3}),
one subtracts the tachyonic pole and adds its residuum to the Higgs pole at $p^2 = m_H^2$.

\section{Perturbative and non-perturbative contributions to the $\rho$-parameter}
\label{sect.rho}

The $\rho$-parameter is  usually defined as the ratio between the neutral and charged current 
 coupling constants at zero momentum transfer
\begin{eqnarray}
\rho = \frac{J_{NC}(0)}{J_{CC}(0)} = \frac{1}{1-\Delta \rho}\,.
\label{eq.58.1}
\end{eqnarray}
$J_{CC}(0)$ is given by the Fermi coupling constant, $G_F$, determined from the $\mu$-decay rate, while
 $J_{NC}(0)$ can be measured in neutrino scattering on electrons or hadrons.
Notice that this definition of the $\rho$-parameter is process dependent since, in general, the radiative corrections
 depend on the hypercharge of the particles involved in the scattering process. 
 However, the leading contributions in the top-quark mass to $\Delta \rho$ are universal. 

At tree-level the $\rho$-parameter is given by $\rho = \frac{M^2_W}{M^2_Z\, c^2_W} = 1$. At the leading order 
 in the top-quark mass, radiative corrections to $\rho$ can be obtained from the wave function renormalization 
 of the unphysical scalars. Let us consider the kinetic terms of  the scalar part of the SM lagrangian.
The UV divergences that show up in radiative corrections can be reabsorbed 
by introducing suitable wave function renormalization constants in the following way
\begin{eqnarray}
\mathcal{L}_{KS} = Z_\phi\,\arrowvert \partial_\mu \phi^- + i \frac{g v}{2}\, W_\mu^- \arrowvert^2 +\frac{Z_\chi}{2}\,
\Big(\partial_\mu \chi +  \frac{g v}{2 c_W}\, Z_\mu \Big)^2+~{\rm other~terms}\,.
\label{eq.58.4}
\end{eqnarray}
The renormalized masses of the gauge bosons are given by 
\begin{eqnarray}
M_W = \sqrt{Z_\phi}\, \frac{g v}{2}\,,~~ 
M_Z = \sqrt{Z_\chi}\, \frac{g v}{2 c_W}\,,~~ {\rm thus}~~ \rho = \frac{Z_\phi}{Z_\chi}\,.
\label{eq.58.5}
\end{eqnarray}

The one-loop wave function renormalization constants of the unphysical scalars have been computed 
 in the previous section, see eqs.(\ref{eq.16n}), (\ref{eq.23n}). By using these results, we immediately 
 get the standard one-loop top contribution to the $\rho$-parameter 
\begin{eqnarray}
\Delta \rho_p = \frac{\alpha_t}{2} + O\big(\alpha_t^2\big) = \frac{\sqrt{2}}{16 \pi^2}\, N_c N_G\, G_F m_t^2 + O\big(G_F^2 m_t^4\big)\,.
\label{eq.58.6}
\end{eqnarray}

Our resummation and subsequent tachyonic subtraction of the scalar propagators allows us to
 give an estimate of the non-perturbative leading top-mass contribution to the $\rho$-parameter. 

 The residuum at the tachyon-pole, $\kappa$, can be viewed as the contribution from the continuous 
 part of the spectrum to the wave function renormalization constants, 
see eqs.(\ref{eq.21com.4}), (\ref{eq.28com.4}) and (\ref{eq.58.4}).
 Therefore,  according to the bMS scheme, one has
\begin{eqnarray}
Z_\chi = \frac{1}{1+\kappa_\chi}\,,~ Z_\phi =\frac{1}{1+\kappa_\phi}~ \Rightarrow~ 
\Delta \rho_c = 1-\frac{1+\kappa_\phi}{1+\kappa_\chi}\,. 
\label{eq.58.7}
\end{eqnarray}
It is worth noticing that the continuous contribution to $\Delta \rho$ is always negative
with this prescription.
 Moreover its absolute value  slowly increases with $\alpha_t$.

It is interesting to compare the bMS approach with the Akhoury scheme. 
In this case we find, by normalizing the spectral densities with a constant factor, the following result:
\begin{eqnarray}
Z_\chi = -\kappa_\chi \,,~ Z_\phi =-\kappa_\phi ~ \Rightarrow~ 
\Delta \rho_c = 1-\frac{\kappa_\chi}{\kappa_\phi}\,. 
\label{eq.58.7bis}
\end{eqnarray}
In the Akhoury scheme the continuous contribution to $\Delta \rho$ is positive, it grows 
 with $\alpha_t$ until it reaches its maximum value, $\Delta \rho_c \simeq 0.038$, for $\alpha_t \simeq 1$
 and eventually it starts decreasing.
We notice that the behaviour in the two schemes is quite different. 
This reflects the uncertainties in the definition of resummation in  improved
perturbation theory. The Akhoury scheme 
appears to be more in agreement with the idea that the improvement of perturbation
theory through the summation of the renormalon chain should act as a cut-off of the theory.
The results of the bMS-like calculation are hard to interpret physically. We notice that
the bMS scheme in this case is not as well motivated as in QCD, where asymptotic freedom acts as an 
additional guiding principle.

\section{Top propagator}
\label{sect.top}

\begin{figure}[h]
\begin{center}
\includegraphics[width=0.9\textwidth]{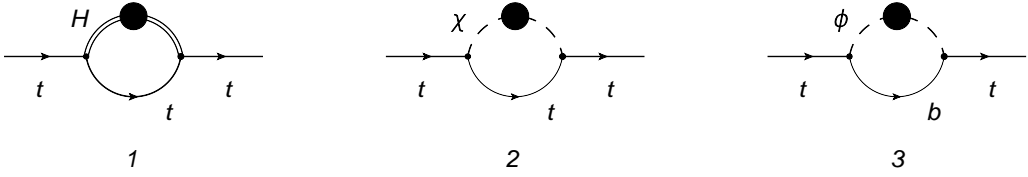}
\end{center}
\caption{Top quark self-energy at one-loop in the gaugeless limit. The lines with a bubble denote a resummed scalar propagator.} 
\label{fig.4}
\end{figure}

In this section we  discuss the one-loop self-energy corrections to the top-quark 
propagator and their renormalization in the on-shell scheme. It turns out that all the contributing graphs  
are of order $O(1)$ in the large $N_G$-limit. We select a gauge invariant subset of self-energy amplitudes by 
 considering the limit of vanishing gauge coupling constants, i.e. $g, g' \to 0$.
 Indeed, in this approximation, which amounts to neglecting the vector boson masses w.r.t. the Higgs boson and the top-quark masses,
 one is left with the Feynman graphs depicted in Fig.~\ref{fig.4}. 

The contribution of graph 1 is given by (for the notation see Appendix~\ref{app.1})
\begin{eqnarray}
i \sqrt{2}\, G_F\, m_t^2\, \Bigg[\big(\slshp + m_t\big)\!
\int_{0}^{+\infty}\!\!\! ds\, \,  \rho_{phy}^H(s) \, B_0\big[p^2,\sqrt{s},m_t\big]
-\slshp \! \int_{0}^{+\infty}\!\!\! ds\, \,  \rho_{phy}^H(s)\, B_1\big[p^2,\sqrt{s},m_t\big]\Bigg]\,, 
\label{eq.topH.1}
\end{eqnarray}
where $\rho_{phy}^H(s) = \delta\big(s-m^2_H\big) + \rho_+^H(s)\,\theta\big(s-4 m^2_t\big)$ is the physical, subtracted 
spectral function. Therefore, one finds a contribution from the Higgs pole which is just the one-loop amplitude and a 
 contribution from the continuous part of the spectrum. 

Some comments are in order. i) Both the scalar and the rank one tensor two point functions in the above equation, i.e. $B_0[p^2,\sqrt{s},m_t]$ 
and $B_1[p^2,\sqrt{s},m_t]$ respectively, have a physical threshold at $p^2 = (\sqrt{s}+m_t)^2$. Since the continuous part of the spectrum 
 starts at $\sqrt{s} = 2 m_t$, the contribution of the latter to the self-energy contains an imaginary part only for $p^2 > 9 m_t^2$. ii) The 
 integral of the imaginary part is convergent being over a compact domain, namely  $4m_t^2 < s < (\sqrt{p^2}-m_t)^2$. 
 iii) The integral over the spectrum of the real part of the two point functions in eq.(\ref{eq.topH.1}) does not converge.         
 In order to see this, we can limit ourselves to the case $p^2 \leq \big(\sqrt{s}-m_t\big)^2$, for the remaining parts of the integrals, 
 if any, are convergent being over a compact domain. By using eqs.(\ref{eq.app.9.4}) and (\ref{eq.app.11.1}), one can show that at the 
leading order in the limit $s \gg p^2,\, m_t^2$,  eq.(\ref{eq.topH.1}) reads 
\begin{eqnarray}
\frac{\sqrt{2}}{16 \pi^2}\, G_F\, m_t^2 
&&\!\!\!\!\!\!\!\! 
\Bigg\{\big(\slshp + m_t\big)\!\int^{+\infty} \!\! ds\, \,  \rho_+^H(s) 
\,\Bigg[\frac{2}{D-4}+1+\log\Big(\frac{s}{\Lambda_B^2}\Big)\Bigg] 
\nonumber\\
&&\!\!\!\!
-\frac{\slshp}{2} \! \int^{+\infty} \!\! ds\, \,  \rho_+^H(s)\, 
\Bigg[\frac{2}{D-4}+\frac{1}{2}+\log\Big(\frac{s}{\Lambda_B^2}\Big)\Bigg]\Bigg\}\,. 
\label{eq.topH.2}
\end{eqnarray}
Both integrals in the above equation are logarithmically divergent.  

The previous considerations suggest that the on-shell renormalization of the top-quark self-energy, besides removing the poles in $D-4$, improves 
 also the behaviour of the integrals over $s$. This is indeed the case. In fact, in the limit where $s \gg p^2, m_t^2$, the subtracted 
 two point functions go to zero as $1/s$ 
\begin{eqnarray}
&&\!\!\!\!\!\!\!\!\!\!\!\!\!\!\!
DB_0\big[p^2,\sqrt{s},m_t\big] := B_0\big[p^2,\sqrt{s},m_t\big]-{\rm Re} \big(B_0\big[p^2 = m^2_t,\sqrt{s},m_t\big]\big) = 
\frac{m^2_t-p^2}{2 s} + O\Big(\frac{1}{s^2}\Big)\,,\nonumber\\
&&\!\!\!\!\!\!\!\!\!\!\!\!\!\!\!
DB_1\big[p^2,\sqrt{s},m_t\big] := 
B_1\big[p^2,\sqrt{s},m_t\big]-{\rm Re} \big(B_1\big[p^2 = m^2_t,\sqrt{s},m_t\big]\big) = \frac{m^2_t-p^2}{3 s} + O\Big(\frac{1}{s^2}\Big)\,.
\label{eq.topH.fun}
\end{eqnarray}
%
This behaviour entails that the integral over the positive part of the spectrum of the subtracted functions in eq.(\ref{eq.topH.fun}) multiplied by 
 $\rho^H_+(s)$ is convergent and can be computed numerically.

We now move to the contribution of the second graph in Fig.~\ref{fig.4}
\begin{eqnarray}
i \sqrt{2}\, G_F\, m_t^2\, \Bigg[\big(\slshp - m_t\big)\!
\int_{0}^{+\infty}\!\!\! ds\, \,  \rho_{phy}^\chi(s) \, B_0\big[p^2,\sqrt{s},m_t\big]
-\slshp \! \int_{0}^{+\infty}\!\!\! ds\, \,  \rho_{phy}^\chi(s)\, B_1\big[p^2,\sqrt{s},m_t\big]\Bigg]\,, 
\label{eq.topc.1}
\end{eqnarray}
where $\rho_{phy}^\chi(s) = \delta\big(s\big) + \rho_+^\chi(s)\,\theta\big(s-4 m^2_t\big)$ is the physical, subtracted 
spectral function. Thus, one finds a contribution from the massless Goldstone pole which is just the one-loop amplitude and a 
 contribution from the continuous part of the spectrum.  The latter coincides, a part for the spectral function and the sign of the 
 mass term, with the continuous contribution to the top-Higgs bubble in eq.(\ref{eq.topH.1}).

Finally, we report here the expression of the third graph in Fig.~\ref{fig.4} in the limit of vanishing bottom mass 
\begin{eqnarray}
2 i \sqrt{2}\, G_F\, m_t^2\, \,\slshp\, \omega_+  \!
\int_{0}^{+\infty}\!\!\! ds\, \,  \rho_{phy}^\phi(s) \, \Big(B_0\big[p^2,\sqrt{s},0\big]
- B_1\big[p^2,\sqrt{s},0\big]\Big)\,, 
\label{eq.topp.1}
\end{eqnarray}
where $\rho_{phy}^\phi(s) = \delta\big(s\big) + \rho_+^\phi(s)\,\theta\big(s- m^2_t\big)$ is the physical, subtracted 
spectral function, while $\omega_+ = \frac{1+\gamma_5}{2}$ is the positive chirality projector. 
Also in this case, one finds a contribution from the massless Goldstone pole which is just the one-loop amplitude and a 
 contribution from the continuous part of the spectrum. 
Arguments similar to those presented for the top-Higgs bubble allow us to conclude that the continuous contribution to the 
 self-energy coming from the bottom-$\phi$ bubble contains an imaginary part for $p^2 > m_t^2$. The on-shell renormalization 
 of the self-energy amplitude in eq.(\ref{eq.topp.1}) guarantees the convergence of the integral over the positive part of 
 the spectrum. Indeed, in the limit where $s \gg p^2,\, m_t^2$, one finds:
\begin{eqnarray}
2\Big(DB_0[p^2,\sqrt{s},0]-DB_1[p^2,\sqrt{s},0]\Big) =
\frac{m^2_t-p^2}{3 s} + O\Big(\frac{1}{s^2}\Big)\,.
\label{eq.topp.2}
\end{eqnarray}

It is convenient to parametrize the top-quark self-energy, $\widehat{\Sigma}_t(p)$, by means of momentum and mass form factors according to the following 
 definition:
\begin{eqnarray}
&&\!\!\!\!\!\!\!\!\!\!\!\!\!\!\!
\widehat{\Sigma}_t(p) = \frac{\sqrt{2}}{16 \pi^2}\, G_F\, m_t^2\, \Big[\Big(a_+^{1l}(p^2) + a^{{\rm c}}_{+}(p^2,\alpha_t)\Big)\, \slshp\,\omega_+  +
\Big(a_-^{1l}(p^2) + a^{{\rm c}}_{-}(p^2,\alpha_t)\Big)\, \slshp\,\omega_- 
\nonumber\\
&&~~~~~~~~~~~~~~~~~~\,\,
+\Big(a_m^{1l}(p^2)+a_m^{{\rm c}}(p^2,\alpha_t)\Big)\, m_t\Big]\,,
\label{eq.topc.2}
\end{eqnarray}
where the coefficients $a_+$, $a_-$ and $a_m$ are given by:
\begin{eqnarray}
&&
a_+^{1l}(p^2) = \frac{16 \pi^2}{-i} \, \Big(DB_0\big[p^2,m_H,m_t\big]-DB_1\big[p^2,m_H,m_t\big]+
DB_0\big[p^2,0,m_t\big]-DB_1\big[p^2,0,m_t\big]\Big) 
\nonumber\\
&& ~~~~~~~~~~
+\frac{32 \pi^2}{-i} \, \Big(DB_0\big[p^2,0,0\big]-DB_1\big[p^2,0,0\big]\Big)\,,
\nonumber\\
&&
a^{{\rm c}}_+(p^2,\alpha_t) = \frac{16 \pi^2}{-i} \, 
\int_{4 m_t^2}^{+\infty}\!\!\! d s\,\, \big(\rho_+^H(s)+\rho_+^\chi(s)\big)
\Big(DB_0\big[p^2,\sqrt{s},m_t\big]-DB_1\big[p^2,\sqrt{s},m_t\big]\Big)
\nonumber\\
&&~~~~~~~~~~~~~~~
+\frac{32 \pi^2}{-i} \, \int_{m_t^2}^{+\infty}\!\!\! d s\,\, \rho_+^\phi(s)
\Big(DB_0\big[p^2,\sqrt{s},0\big]-DB_1\big[p^2,\sqrt{s},0\big]\Big)\,,
\nonumber\\
&&
a_-^{1l}(p^2) = \frac{16 \pi^2}{-i} \, \Big(DB_0\big[p^2,m_H,m_t\big]-DB_1\big[p^2,m_H,m_t\big]+
DB_0\big[p^2,0,m_t\big]-DB_1\big[p^2,0,m_t\big]\Big)\,, 
\nonumber\\
&&
a^{{\rm c}}_-(p^2,\alpha_t) = \frac{16 \pi^2}{-i} \,
\int_{4 m_t^2}^{+\infty}\!\!\! d s\,\, \big(\rho_+^H(s)+\rho_+^\chi(s)\big)
\Big(DB_0\big[p^2,\sqrt{s},m_t\big]-DB_1\big[p^2,\sqrt{s},m_t\big]\Big)\,,
\nonumber\\
&&
a_m^{1l}(p^2) = \frac{16 \pi^2}{-i} \, \Big(DB_0\big[p^2,m_H,m_t\big]-
DB_0\big[p^2,0,m_t\big]\Big)\,,
\nonumber\\
&&
a_m^{{\rm c}}(p^2,\alpha_t) = \frac{16 \pi^2}{-i} \,
\int_{4 m_t^2}^{+\infty}\!\!\! d s\,\, \big(\rho_+^H(s)-\rho_+^\chi(s)\big)
\, DB_0\big[p^2,\sqrt{s},m_t\big]\,.
\label{eq.topc.3}
\end{eqnarray}

\begin{figure}[p]
\begin{center}
\includegraphics[width=0.45\textwidth]{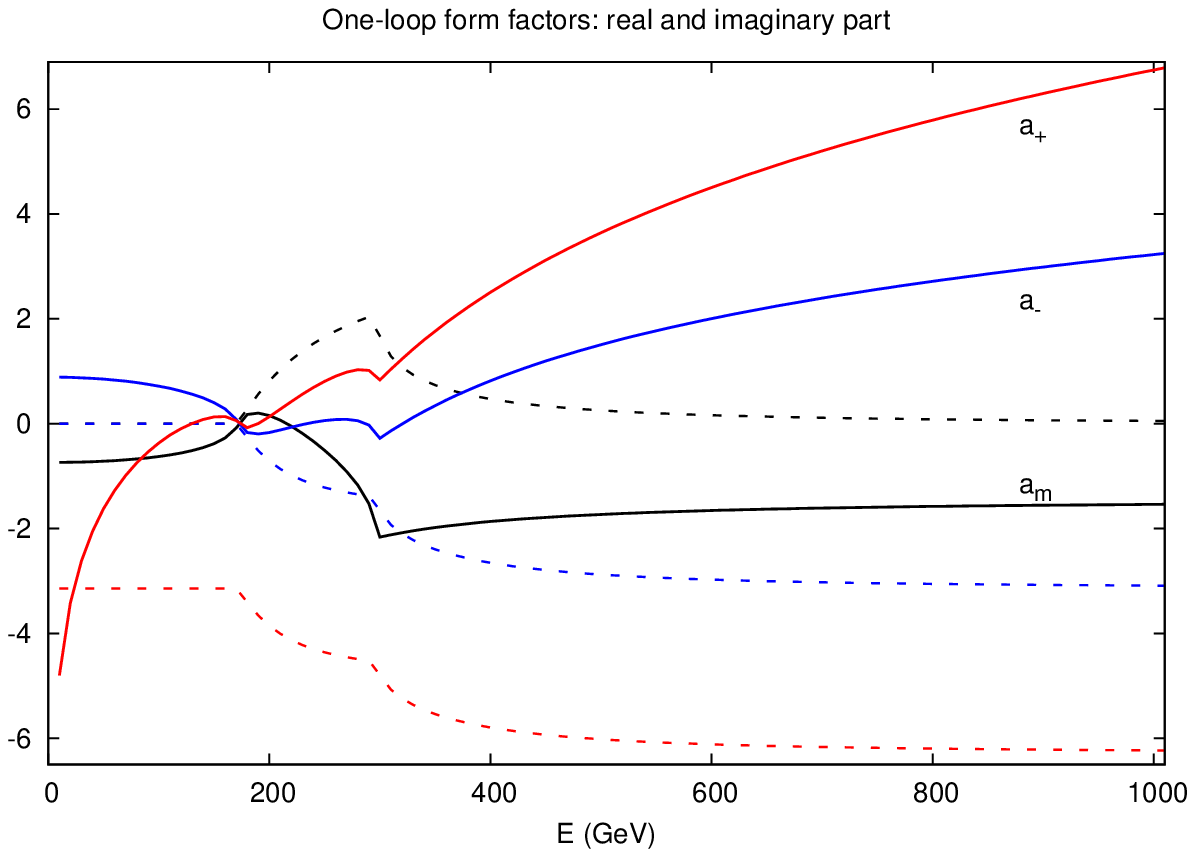}
\includegraphics[width=0.45\textwidth]{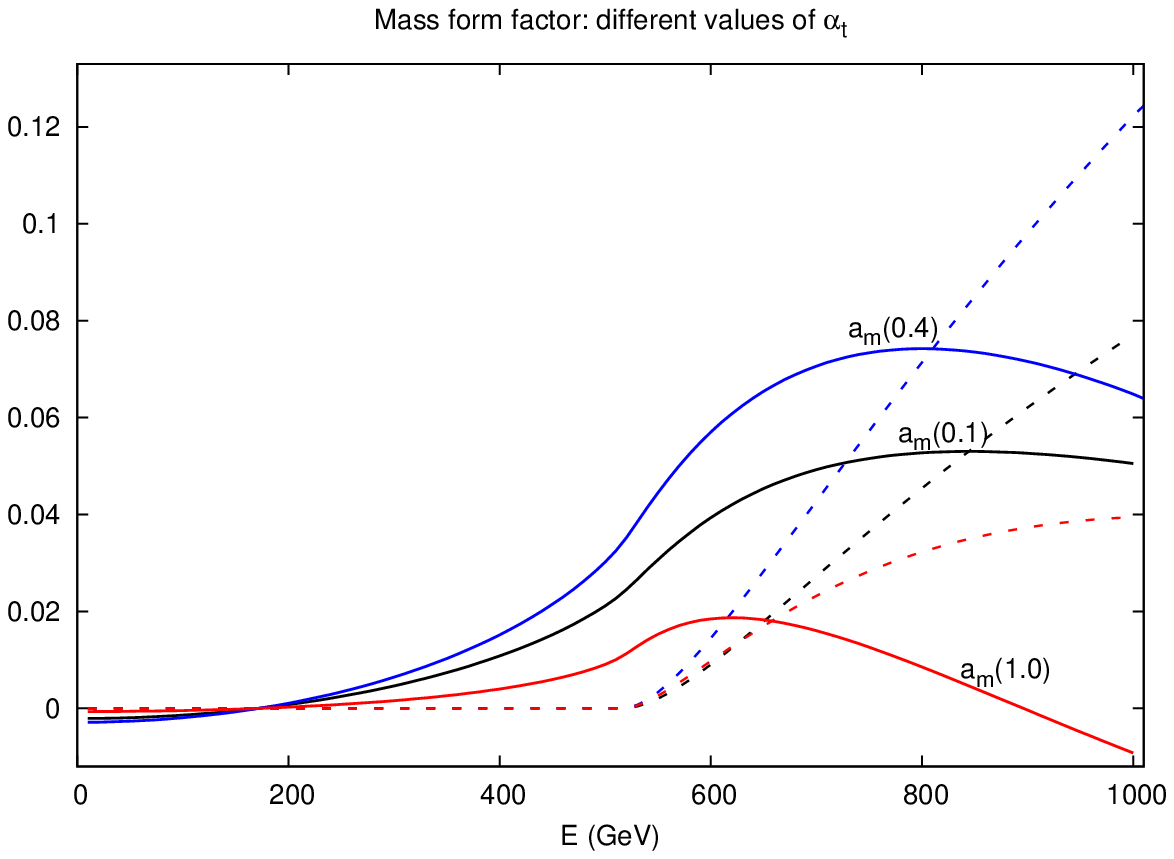}
\end{center}
\caption{Left panel: real (solid lines) and imaginary (dashed lines, same colours) parts of the one-loop form factors. 
Right panel: real (solid lines) and imaginary (dashed lines) parts of the continuous contribution to the mass form factor 
for three values of $\alpha_t$.}
\label{fig.malpha}
\end{figure}

\begin{figure}[p]
\begin{center}
\includegraphics[width=0.45\textwidth]{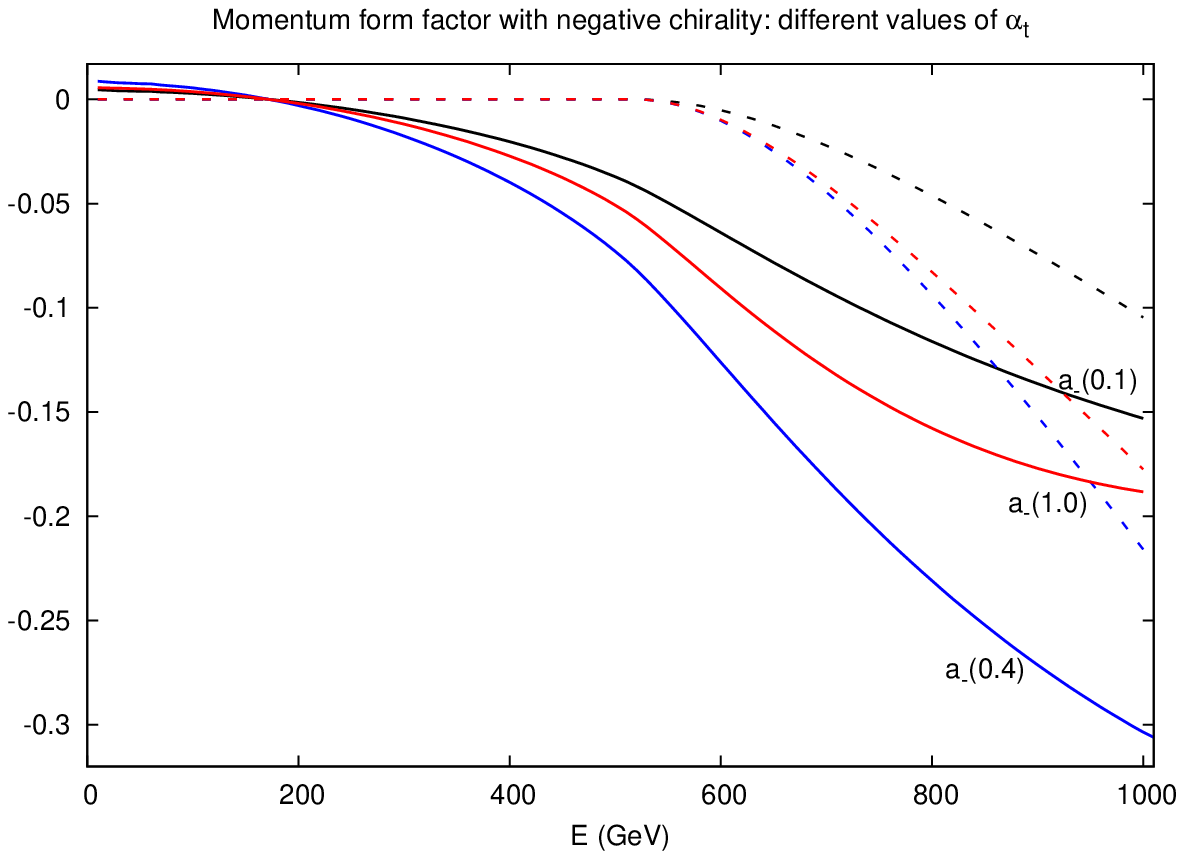}
\includegraphics[width=0.45\textwidth]{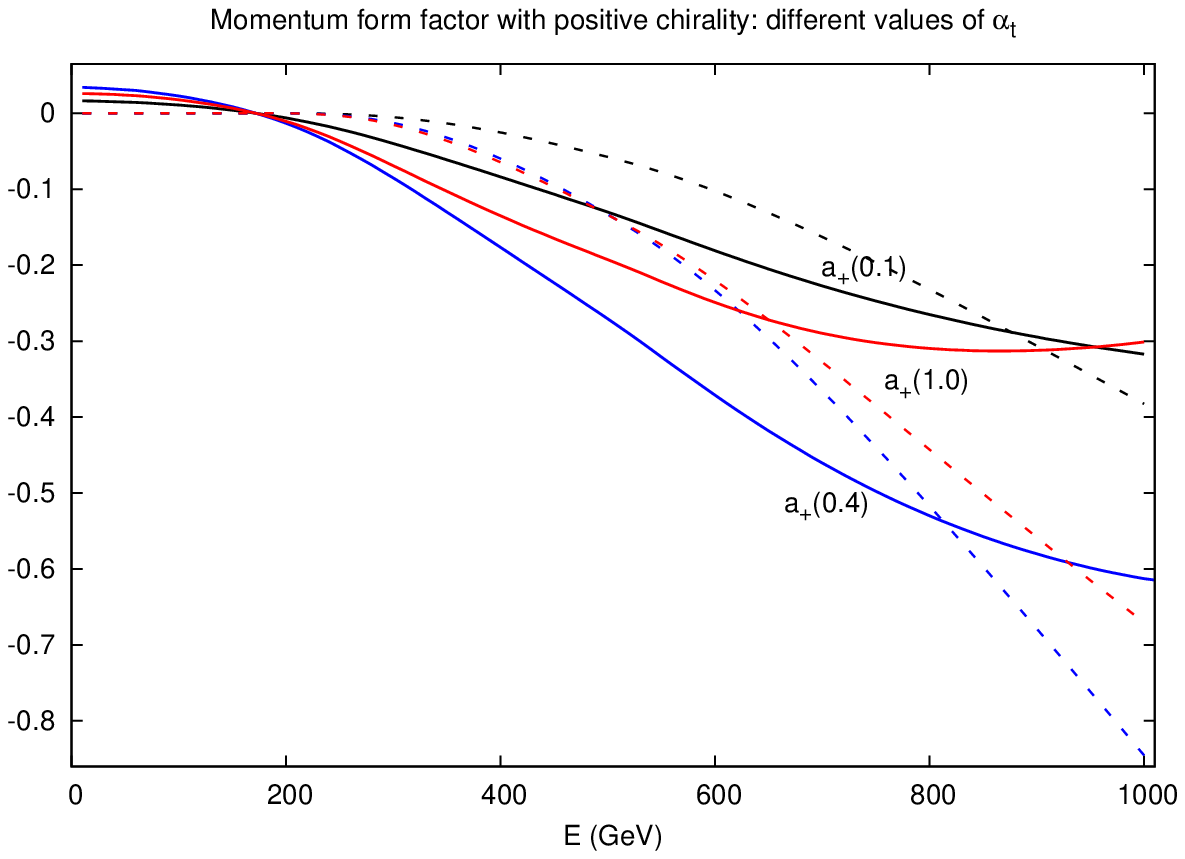}
\end{center}
\caption{Left panel: real (solid lines) and imaginary (dashed lines, same colours) parts of the continuous contribution to the momentum 
form factor with negative chirality for three values of $\alpha_t$. Right panel: same plot, but with positive chirality. }
\label{fig.palpha}
\end{figure}

\begin{figure}[p]
\begin{center}
\includegraphics[width=0.45\textwidth]{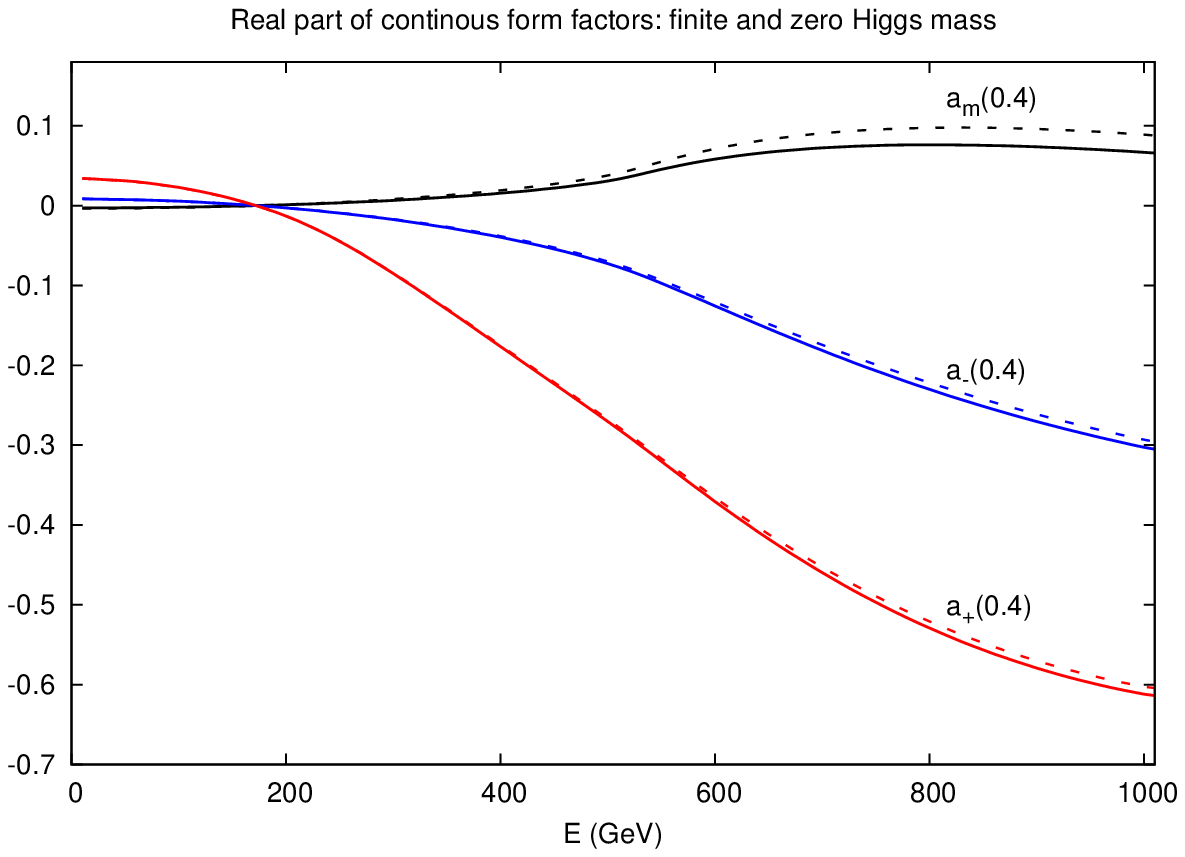}
\includegraphics[width=0.45\textwidth]{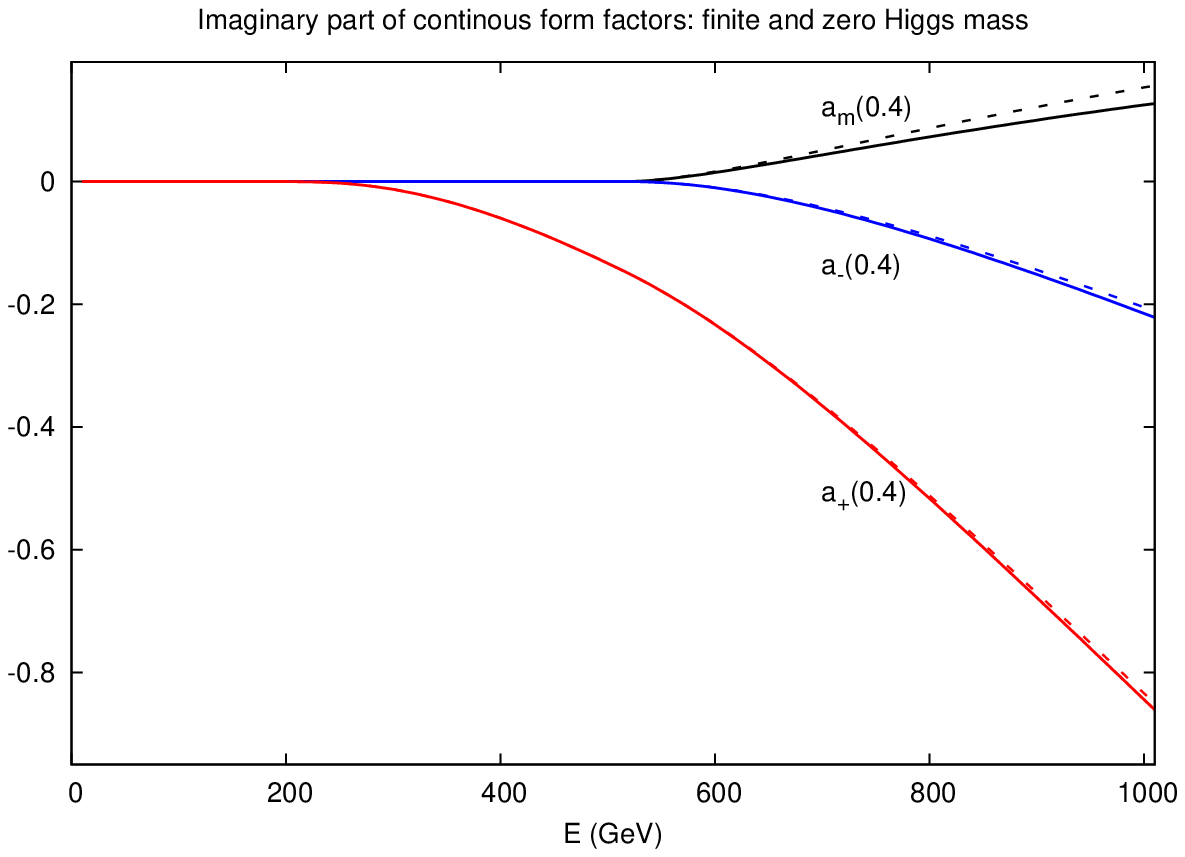}
\end{center}
\caption{Left panel: real parts of the continuous contribution to the form factors computed with $m_H=125\,$ GeV (solid lines) and 
with $m_H = 0\,$ GeV (dashed lines).
Right panel: same plot with the imaginary parts}
\label{fig.H0}
\end{figure}

In Fig.~\ref{fig.malpha} (left panel) we plot the real (solid lines) and imaginary (dashed lines) parts of the one-loop form factors as functions 
 of the external momentum, $\sqrt{p^2}$. The presence of two thresholds at $\sqrt{p^2} = m_t \simeq 170\,$ GeV and  
 $\sqrt{p^2} = m_t+m_H \simeq 300\,$ GeV, is clearly distinguishable. In the right panel of the same figure we show the behaviour of the 
 real (solid lines) and imaginary (dashed lines) parts of the continuous contribution to the mass form factor for 
 different values of the coupling constant $\alpha_t$. The real part of $a_m^c(p^2,\alpha_t)$ as a function of the external momentum has a maximum 
 that depends on $\alpha_t$, though it is always located beyond the threshold at $\sqrt{p^2} = 3 m_t = 516\,$ GeV. The impact of the continuous 
 part of the spectrum on the mass form factor is negligible ($< 5\%$) over the whole range of momentum considered.   
In Fig.~\ref{fig.palpha} the continuous contribution to the real (solid lines) and imaginary (dashed lines) parts of the momentum form factors 
with negative (left panel) and positive (right panel) chirality  is plotted as a function of the external momentum for different values of $\alpha_t$.  
 It turns out that in this case the continuous contribution to the momentum form factors can be a sizable fraction ($5\% - 10\%$) of the corresponding 
 one-loop contribution for $\sqrt{p^2} > 500\,$ GeV. A detailed inspection shows that the effects of the continuous part of the spectrum on all form factors 
 grow with $\alpha_t$ until they reach their maximum for $\alpha_t \simeq 0.4$ and then they decrease.    
Finally, in Fig.~\ref{fig.H0} we present a comparison between the real (left panel) and the imaginary (right panel) 
 parts of the continuous contribution to the 
 form factors computed with a finite Higgs mass (solid lines), namely $m_H = 125\,$ Gev, and in the approximation of vanishing Higgs mass (dashed lines).
 A fixed value of the coupling constant, $\alpha_t = 0.4$, has been used to compute all the form factors. One can see that the impact 
of a finite Higgs mass is small ($< 5\%$) on the momentum form factors, but can be quite big 
 (around $20\% - 25 \%$ for $\sqrt{p^2} < 700\,$ GeV and bigger than $30 \%$ 
 for $\sqrt{p^2} > 800\,$ GeV) on the mass form factor.

%

\subsection{Complex pole of the top propagator}
\label{sect.wid}

In this subsection we compute the complex pole of the Dyson resummed top propagator and extract from it the width of the top-quark.  
Moreover, the impact on this latter quantity of the continuous part of the spectrum is estimated.

The Dyson resummed top propagator,
\begin{eqnarray}
\Delta_t(p) = \frac{i}{\slshp -m_t -\widehat{\Sigma}_t(p)+i \epsilon} 
\label{eq.wid.2}
\end{eqnarray}
can be cast in the following form
\begin{eqnarray}
\Delta_t(p) = \frac{i}{D(p^2)}\, \left[\frac{\slshp\, \omega_+}{1-\tilde{a}_-(p^2,\alpha_t)}+\frac{\slshp\, \omega_-}{1-\tilde{a}_+(p^2,\alpha_t)}
+ \frac{m_t\big(1+\tilde{a}_m(p^2,\alpha_t)\big)}{\big(1-\tilde{a}_-(p^2,\alpha_t)\big)\big(1-\tilde{a}_+(p^2,\alpha_t)\big)}\right]\,, 
\label{eq.wid.3}
\end{eqnarray}
where
\begin{eqnarray}
D(p^2) = p^2-m^2_t\, \frac{\big(1+\tilde{a}_m(p^2,\alpha_t)\big)^2}{\big(1-\tilde{a}_-(p^2,\alpha_t)\big)\big(1-\tilde{a}_+(p^2,\alpha_t)\big)}\,.
\label{eq.wid.4}
\end{eqnarray} 
The form factors appearing in the above equations are given by the rescaled sum of the one-loop and the continuous form factors 
given in eq.(\ref{eq.topc.3}), i.e.  
$$\tilde{a}_i(p^2,\alpha_t) = \frac{\sqrt{2}}{16 \pi^2}\, G_F m_t^2 \,\big[a^{1l}_i(p^2) + a^{c}_i(p^2,\alpha_t)\big]\,,~~ i = +,-,m\,.$$ 

The pole of the resummed propagator, $s$, which in general is a complex quantity, can be obtained by solving numerically the following 
 equation
\begin{eqnarray}
\frac{s}{m_t^2} = \frac{\big(1+\tilde{a}_m(s,\alpha_t)\big)^2}{\big(1-\tilde{a}_-(s,\alpha_t)\big)
\big(1-\tilde{a}_+(s,\alpha_t)\big)}\,.
\label{eq.wid.5}
\end{eqnarray} 
We parametrize the complex pole $s$ in the following way
\begin{eqnarray}
\sqrt{s} = M_t - \frac{i}{2}\, \Gamma_t\,,
\label{eq.wid.8}
\end{eqnarray} 
where $M_t$ is the physical top mass and $\Gamma_t$ its width. 
In this way, neglecting quadratic corrections in the width (narrow width approximation), one finds $s = M_t^2 -i M_t \Gamma_t$. 

In Tab.~\ref{tab.1} we show the width of the top-quark, $\Gamma_t$, as a function of its physical mass, $M_t$.
In particular, we compare results obtained by taking into account one-loop corrections to the top propagator only, with those
 where the contribution coming from the continuous part of the spectrum has been added. It turns out that the impact of non-perturbative
 corrections on the top width is small ($< 7 \%$) for $M_t < 1\,$ TeV, but becomes quite sizable ($> 15 \%$) for heavier masses, typically 
 above $1.5\,$ TeV. Finally, our results show that the presence of a light Higgs boson, with mass $m_H = 125\,$ GeV, affects the width of  
 the top-quark significantly ($> 10 \%$) only if the latter is light, $M_t < 500\,$ Gev.   
  
%
\begin{table}[h]
\begin{center}
\begin{tabular}{|c|c|c|c|c|}
\hline
$M_t$ & $\Gamma_t^{1l}(m_H = 125)$ & $\Gamma^{1l+cont.}_t(m_H = 125)$ & $\Gamma_t^{1l}(m_H=0)$ & $\Gamma_t^{1l+cont.}(m_H=0)$ \\
\hline
172 & 1.6666 & 1.6666 & 1.6670 & 1.6670 \\
\hline
200 & 1.8286 & 1.8286 & 3.7368 & 3.7367 \\
\hline
300 & 2.8969 & 2.9077 & 8.2406 & 8.2510 \\
\hline
400 & 9.4270 & 9.4977 & 11.6188 & 11.6894 \\
\hline
500 & 13.2124 & 13.4136 & 14.7659 & 14.9677 \\
\hline
600 & 16.6049 & 16.9956 & 17.8448 & 18.2295 \\ 
\hline
700 & 19.8499 & 20.5433 & 20.8998 & 21.5559 \\
\hline
800 & 23.0236 & 24.1733 & 23.9460 & 25.0080 \\
\hline 
900 & 26.1581 & 27.9215 & 26.9893 & 28.6014 \\
\hline
1000 & 29.2691 & 31.8017 & 30.0325 & 32.3400 \\
\hline
1100 & 32.3649 & 35.8185 & 33.0767 & 36.2235 \\
\hline
1200 & 35.4509 & 39.9722 & 36.1224 & 40.2493 \\
\hline
1300 & 38.5300 & 44.2607 & 39.1699 & 44.4138 \\
\hline
1400 & 41.6046 & 48.6807 & 42.2194 & 48.7126 \\
\hline
1500 & 44.6758 & 53.2280 & 45.2707 & 53.1409 \\
\hline
1600 & 47.7449 & 57.8979 & 48.3240 & 57.6941 \\
\hline
1700 & 50.8124 & 62.6856 & 51.3791 & 62.3670 \\
\hline
1800 & 53.8790 & 67.5863 & 54.4360 & 67.1549 \\
\hline
1900 & 56.9449 & 72.5950 & 57.4945 & 72.0528 \\
\hline
2000 & 60.0106 & 77.7069 & 60.5548 & 77.0560 \\
\hline
\end{tabular}
\caption{Top quark's width expressed in GeV.}
\label{tab.1}
\end{center}
\end{table}
%

\section{Bottom propagator}
\label{sect.bot}

In this section we  discuss the one-loop self-energy corrections to the bottom-quark 
propagator and their renormalization in the on-shell scheme. It turns out that all the contributing graphs  
are of order $O(1)$ in the large-$N_G$ limit. We select a gauge invariant subset of self-energy amplitudes by 
considering the limit where $g, g' \to 0$. The resulting graphs can be obtained from those depicted 
in Fig.~\ref{fig.4} by substituting a top propagator with a bottom one and vice versa. 

The contribution of graphs 1 and 2 to the bottom self-energy, $\Sigma_b(p)$, is proportional 
to $G_F m_b^2$ and thus it can be neglected. The expression of the third graph is given by
\begin{eqnarray}
2 i \sqrt{2}\, G_F\, m_t^2\, \,\slshp\, \omega_-  \!
\int_{0}^{+\infty}\!\!\! ds\, \,  \rho_{phy}^\phi(s) \, \Big(B_0\big[p^2,\sqrt{s},m_t\big]
- B_1\big[p^2,\sqrt{s},m_t\big]\Big)\,. 
\label{eq.bot.1}
\end{eqnarray}
In the above equation one finds a contribution from the massless Goldstone pole which is just the one-loop amplitude and 
 a contribution from the continuous part of the spectrum. The latter has a physical threshold, above which an imaginary part shows up, at 
 $p^2 = 4 m_t^2$ due to the fact that the continuous part of the spectrum starts at $\sqrt{s} = m_t$. 
Moreover, the on-shell renormalization of the bottom self-energy in eq.(\ref{eq.bot.1}) guarantees the convergence 
 of the integral over the positive part of the spectrum.

It is convenient to parametrize the bottom-quark self-energy, $\widehat{\Sigma}_b(p)$, by means of a momentum form factor according to the following 
 definition:
\begin{eqnarray}
\widehat{\Sigma}_b(p) = \frac{\sqrt{2}}{16 \pi^2}\, G_F\, m_t^2\, \Big(b_{-}^{1l}(p^2) + b^{{\rm c}}_{-}(p^2,\alpha_t)\Big)\, \slshp\,\omega_-  
\label{eq.bot.2}
\end{eqnarray}
where the coefficients are given by:
\begin{eqnarray}
&&
b_-^{1l}(p^2) = \frac{32 \pi^2}{-i} \, \Big(DB_0\big[p^2,0,m_t\big]-DB_1\big[p^2,0,m_t\big]\Big)
\nonumber\\
&&
b^{{\rm c}}_-(p^2,\alpha_t) = \frac{32 \pi^2}{-i} \, 
\int_{m_t^2}^{+\infty}\!\!\! d s\,\, \rho_+^\phi(s)
\Big(DB_0\big[p^2,\sqrt{s},m_t\big]-DB_1\big[p^2,\sqrt{s},m_t\big]\Big)\,.
\label{eq.bot.3}
\end{eqnarray}

\begin{figure}[h]
\begin{center}
\includegraphics[width=0.45\textwidth]{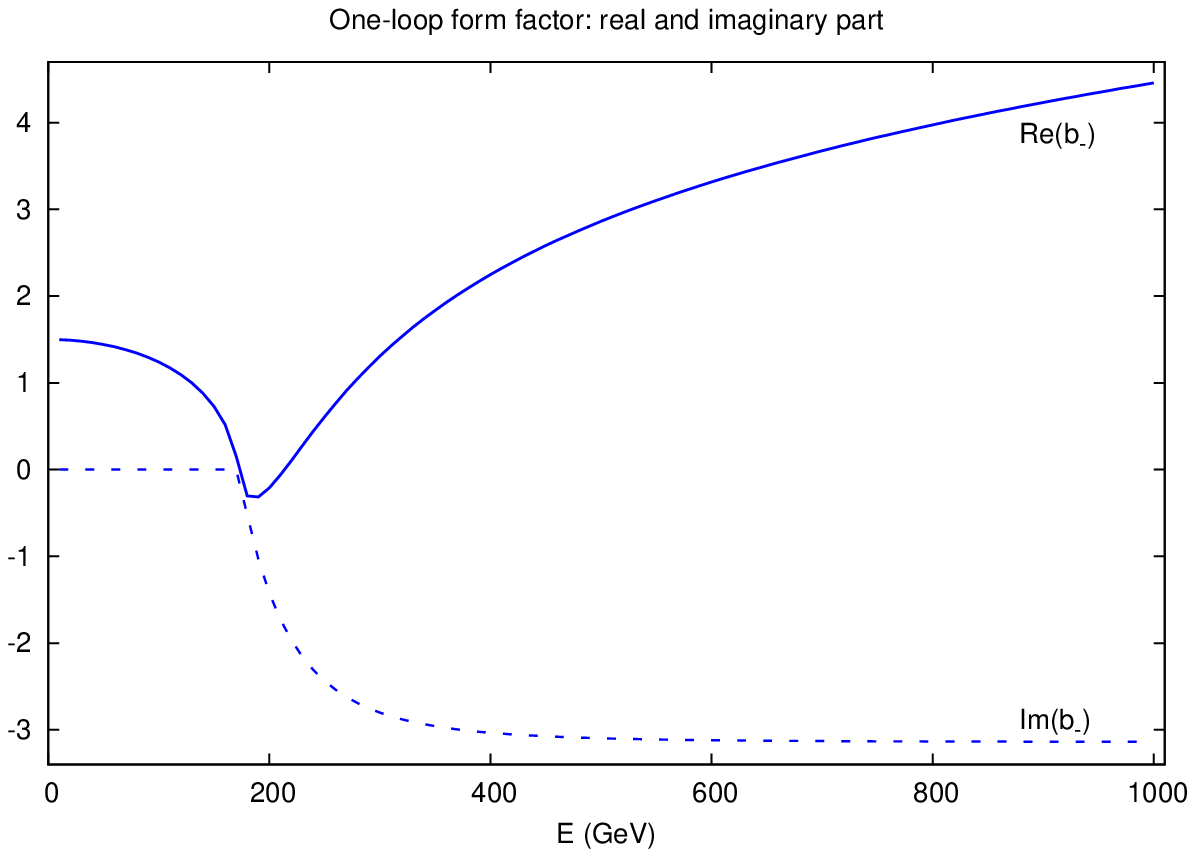}
\includegraphics[width=0.45\textwidth]{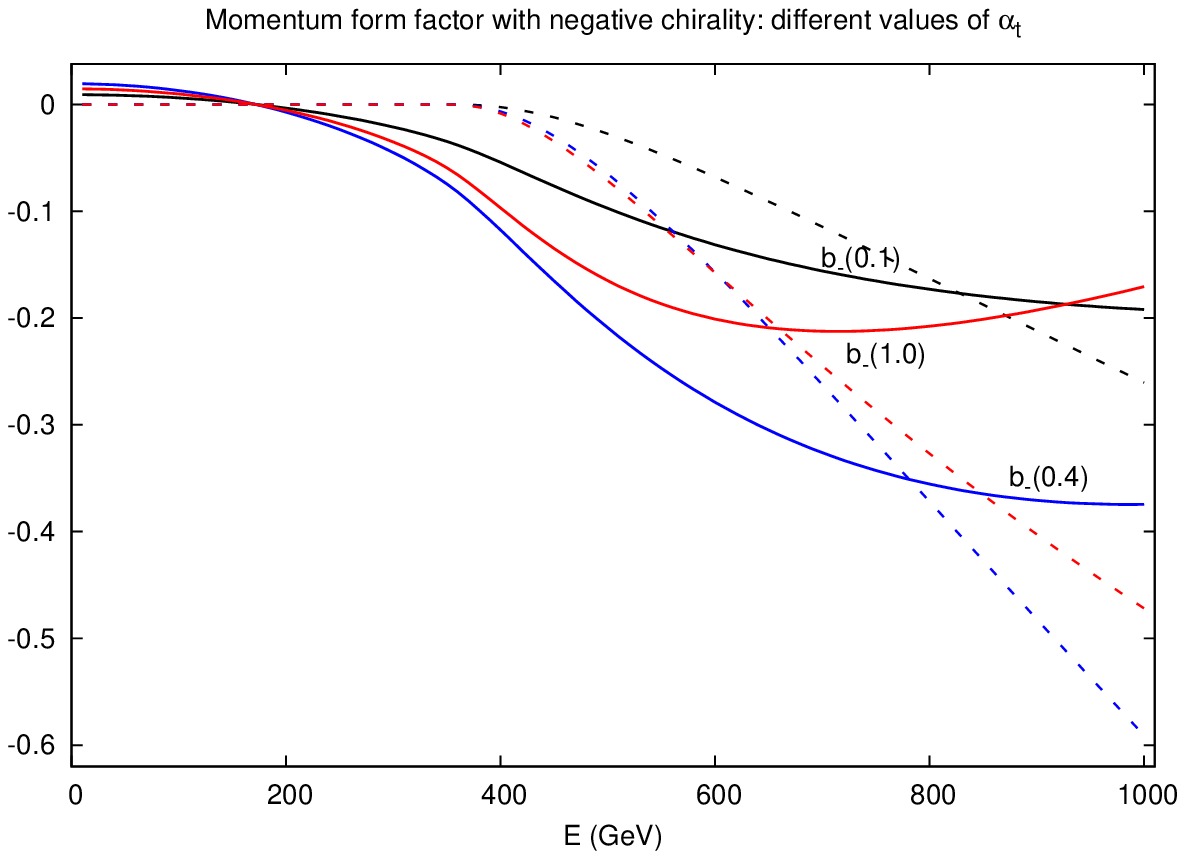}
\end{center}
\caption{Left panel: real (solid line) and imaginary (dashed line) part of the one-loop form factor. Right panel: real (solid lines) 
and imaginary (dashed lines, same colours) parts of the continuous contribution to the momentum form factor for three values of $\alpha_t$.}
\label{fig.bpmalpha}
\end{figure} 

In the left panel of Fig.~\ref{fig.bpmalpha} we show the real (solid line) and imaginary (dashed line) part of the 
 one-loop momentum form factor of the bottom propagator. In the right panel of the same figure we plot the 
 real (solid lines) and imaginary (dashed lines) parts of the continuous contribution to $b_-$ for three different values of $\alpha_t$. 
 It turns out that the impact of the continuous part of 
 the spectrum on the momentum form factor is not negligible being about $5 \% - 10 \%$ of the one-loop 
 contribution for $\sqrt{p^2} > 500\,$ GeV. Finally, also in the case of the bottom propagator the effect of the continuous part 
 of the spectrum on the form factor reaches its maximum for $\alpha_t \simeq 0.4$.

\section{Conclusion}
\label{sect.con}

The question whether the presence of a Landau pole, i.e. a tachyon-pole in
the propagator, signifies the breakdown of a theory or whether it is an artifact
of perturbation theory is a difficult question. It has been with us
for 60 years and one still cannot claim that the problem is solved.
Within QCD analytic perturbation theory appears to give fundamentally
correct results, however here one uses asymptotic freedom as an essential
subsidiary principle. In this paper we attempted to resum perturbation theory
in a similar method by at least first subtracting the tachyon
and subsequently calculate with the corrected propagator. We showed that such calculations are
feasible in the electroweak sector. We focused on effects of a heavy top-quark which
simplifies the discussion considerably, since the problems then appear in one place in
the theory only and can be studied in isolation. Also the heavy top effects are the largest
in the SM, however perturbation theory is surely sufficient for the physical top-quark mass.
Nonetheless the calculations are important for possible effects of a (very unlikely)
fourth family or effects from fermion doubles, when one tries to take the continuum limit
of a lattice action.

Lacking the extra input from asymptotic freedom, one needs  new principles in order to constrain
the uncertainties coming from  non-perturbative effects. Following a previous paper
in Higgs physics, we introduced the Akhoury scheme, which appears to give sensible
results. The scheme was motivated by principles from renormalization theory like
the normalization of the spectral density integral. An attempt to generalize analytic perturbation
theory gave quite different results, that do not look very meaningful. However, since rigorous
principles constraining the treatment of non-perturbative uncertainties are missing in the electroweak
case, we cannot come to a definite conclusion. It would be very useful if cut-off effects
could be studied in an entirely different non-perturbative scheme, for instance with a lattice
lagrangian. However at the moment it appears unclear how one should put a chiral model
with a heavy top-quark and a massless bottom-quark on the lattice. In particular, the fermion doubling
problem will complicate things here due to the lack of (perturbative) decoupling of
heavy fermion doubles.

\section*{Acknowledgements}
This work is supported by the DFG project "{(Nicht)-perturbative Quantenfeldtheorie\,}".

\appendix
\section{Tree-level lagrangian}
\label{app.2}

For completeness we report here the lagrangian that has been used in the computations.
We have adopted a Landau gauge-fixing in order to have massless would-be Goldstone bosons.
We omit the vertices with three and four gauge fields, the ghost
part of the lagrangian and all the fermion fields except the top and bottom-quarks since they do not play a role in
our computation. The bilinear part of the lagrangian is given by
\begin{eqnarray}
\mathfrak{L}_{bil} \!\!\!&=&\!\!\! W_\mu^+\, g^{\mu\nu}\big(\square +M^2_W\big) W_\nu^- + \frac{1}{2} \,
Z_\mu \,g^{\mu\nu}\big(\square +M^2_Z\big) Z_\nu +\frac{1}{2}\, A_\mu\, g^{\mu\nu}\square A_\nu \nonumber\\
&&\!\!\!\!\! - \, \phi^+ \,\square  \phi^-
- \frac{1}{2} \, H \,\big(\square +m^2_H \big) H - \frac{1}{2} \, \chi \,\square \chi
 \nonumber\\
&&\!\!\!\!
 + \sum_{k=1}^{N_G} \Big[ \bar{t}_k \big( i\, \slshpart - m_t \big) t_k +\bar{b}_k \, i\, \slshpart b_k \Big]\,,
\label{eq.lag.1}
\end{eqnarray}
where $M_W = \frac{1}{2}\, g\, v$, $M_Z = \frac{M_W}{c_w}$, $m_H = \frac{1}{\sqrt{2}}\, \sqrt{\lambda}\,v$ and
 $m_t = \frac{1}{\sqrt{2}}\, y_t\, v$ and $A_\mu$ is the photon field.

The trilinear part of the lagrangian is given by
\begin{eqnarray}
\!\!\!\!\!\!\!\!\!\!\!\!
\mathfrak{L}_{tri} \!\!\!&=&\!\!\! \frac{i}{2}\, g\, W_\mu^+ \Big[ \phi^- \big(\partial^\mu H +i\, \partial^\mu
\chi\big)
- \big(H+i\, \chi\big) \partial^\mu \phi^-\Big] \nonumber\\
&&\!\!\!\!
+\frac{i}{2}\, g\, W_\mu^- \Big[ \big(H - i\, \chi\big) \partial^\mu \phi^+
-\phi^+ \big(\partial^\mu H -i\, \partial^\mu \chi\big)\Big]
 \nonumber\\
&&\!\!\!\!
+\frac{i}{2}\, g\, Z_\mu \Big( \phi^- \partial^\mu \phi^+
-\phi^+ \partial^\mu \phi^- + i\, \chi \, \partial^\mu H -i\, H \, \partial^\mu \chi \Big)
 \nonumber\\
&&\!\!\!\!
+\frac{1}{4}\, g^2 \, v\, H\, \big(2 W^+ \cdot W^- +Z^2\big) -\frac{1}{4}\, \lambda\, v \,
\big(H^3+ H\,\chi^2 +2 H\, \phi^+ \phi^-\big)
 \nonumber\\
&&\!\!\!\!
+ \frac{1}{\sqrt{2}}\, g\,\sum_{k=1}^{N_G} \Big( \bar{t}_k\, \slshW^{~\, +}\, \omega_-\, b_k +
 \bar{b}_k\, \slshW^{~\, -}\, \omega_-\, t_k\Big)
+ \frac{1}{2}\, g \, \sum_{k=1}^{N_G} \Big( \bar{t}_k\, \slshZ \,\,\, \omega_-\, t_k -
\bar{b}_k\, \slshZ \,\,\, \omega_-\, b_k\Big)
 \nonumber\\
&&\!\!\!\!
+y_t\, \sum_{k=1}^{N_G} \Big(\phi^+ \, \bar{t}_k\,  \omega_-\, b_k + \phi^-\,
\bar{b}_k\, \omega_+\, t_k \Big)
-\frac{1}{\sqrt{2}}\, y_t\, \sum_{k=1}^{N_G} \Big(H\, \bar{t}_k\, t_k - i\, \chi\, \bar{t}_k\, \gamma_5\, t_k
\Big)\,.
\label{eq.lag.2}
\end{eqnarray}

Finally the quadrilinear part of the lagrangian is given by
\begin{eqnarray}
\mathfrak{L}_{quad} \!\!\!&=&\!\!\! \frac{1}{8}\, g^2\, \Big(2 W^+ \cdot W^- +Z^2\Big)
\Big(2 \phi^+\, \phi^- + H^2 + \chi^2\Big)\nonumber\\
&&\!\!\!\!\! -\frac{1}{16}\, \lambda \Big[4 \big(\phi^+\, \phi^-\big)^2 + H^4 + \chi^4 + 4 \phi^+\,
\phi^-\, H^2 + 4 \phi^+\, \phi^-\, \chi^2 +  2 H^2\, \chi^2 \Big]\,.
\label{eq.lag.3}
\end{eqnarray}
We remark that in the above equations all the mass parameters, the coupling constants and the fields are bare
quantity, eventhough a subscript $"0"$ has not been added in order to avoid a cumbersome notation.

\section{One-loop scalar integrals}
\label{app.1}

We collect in this Appendix some useful formulas that have been used in this work.
We denote with $\mu$ the mass scale introduced with dimensional regularization.
Given the following definitions (with $n$  a positive integer)
\begin{eqnarray}
&&
A_0^{(n)}[m] = \int \!\! \frac{d^Dq}{(2\pi)^D}\, \frac{\mu^{(4-D)}}
{\big(q^2-m^2 + i\epsilon\big)^n}\,\,,
\nonumber\\
&&
B_0[p^2,m,M] = \int \!\! \frac{d^D q}{(2\pi)^D}\,\, \frac{\mu^{(4-D)}}{\big(q^2-m^2+i\epsilon\big)
 \big[(q-p)^2-M^2+i\epsilon\big]}\,\, ,
\label{eq.app.1}
\end{eqnarray}
it is straightforward to derive the explicit expression for the one-point functions $A^{(n)}_0$
\begin{eqnarray}
&&
A^{(1)}_0[m]\equiv A_0[m] = \frac{i}{(4\pi)^2}\, m^2 \Big[
\frac{2}{4-D}-1-\log\Big(\frac{m^2}{\Lambda_B^2}\Big)\Big]\,,
\nonumber\\
&&
A^{(2)}_0[m] = \frac{i}{(4\pi)^2}\, \Big[
\frac{2}{4-D}-2-\log\Big(\frac{m^2}{\Lambda_B^2}\Big)\Big]\,,
\nonumber\\
&&
A_0^{(n)}[m] = \frac{(-)^n i}{(4\pi)^2}\, \frac{1}{\big(n-2\big) \big(n-1\big)}\, \frac{1}{m^{2(n-2)}}\,,~
 {\rm for}~ n > 2\,,
\label{eq.app.2}
\end{eqnarray}
where $\Lambda^2_B = 4 \pi\, \mu^2\, \exp\big(2-\gamma_{_E}\big)$.

Before dealing with the complete expression of the scalar two-point function $B_0$, 
 we consider some special cases. We start with the simplest case, the one with 
two massless particles.
\begin{eqnarray}
B_0[p^2,0,0] = \frac{i}{(4\pi)^2}\Big[
\frac{2}{4-D}-\log\Big(-\frac{p^2}{\Lambda_B^2}- i \epsilon\Big)\Big] \,.
\label{eq.app.7}
\end{eqnarray}
The two-point function in the case of one massive particle and one massless particle is given by
\begin{eqnarray}
B_0[p^2,m,0] = \frac{i}{(4\pi)^2}\Big[
\frac{2}{4-D}-\log\Big(\frac{m^2}{\Lambda_B^2}\Big)-
\Big(1-\frac{m^2}{p^2}\Big)\log\Big(1-\frac{p^2}{m^2}\Big)\Big] \,.
\label{eq.app.5}
\end{eqnarray}
The above expression is valid in the kinematical region $p^2 < m^2$, while above the
 production threshold, $p^2 > m^2$,  we have
\begin{eqnarray}
\!\!\!\!\!
B_0[p^2,m,0] = \frac{i}{(4\pi)^2}\Big[
\frac{2}{4-D}-\log\Big(\frac{m^2}{\Lambda_B^2}\Big)-
\Big(1-\frac{m^2}{p^2}\Big)\log\Big(\frac{p^2}{m^2}-1\Big)+i\,\pi\Big(1-\frac{m^2}{p^2}\Big)\Big] .
\label{eq.app.5.1}
\end{eqnarray}
By using eq.(\ref{eq.app.5}) it is easy to prove that
\begin{eqnarray}
\lim_{p^2\to 0} B_0[p^2,m,0] =\frac{1}{m^2}\, A_0[m] = \frac{i}{(4\pi)^2} \Big[
\frac{2}{4-D}-1-\log\Big(\frac{m^2}{\Lambda_B^2}\Big)\Big]\,.
\label{eq.app.5.2}
\end{eqnarray}

The two-point function in the case of two massive particles with equal mass below the production threshold, $p^2 < 4 m^2$, reads
\begin{eqnarray}
B_0[p^2,m,m] = \frac{i}{(4\pi)^2}\Bigg[
\frac{2}{4-D}-\log\Big(\frac{m^2}{\Lambda_B^2}\Big)-
2\,\sqrt{-\Delta}\,\arctan\Bigg(\frac{1}{\sqrt{-\Delta}}\Bigg)\Bigg] \,,
\label{eq.app.9}
\end{eqnarray}
where we have introduced the shorthand notation $\Delta = 1-\frac{4m^2}{p^2}$ .\\
The two-point function with equal masses for $p^2 > 4 m^2$ is given by
\begin{eqnarray}
\!\!\!\!\!\!\!
B_0[p^2,m,m] = \frac{i}{(4\pi)^2}\Bigg\{
\frac{2}{4-D}-\log\Big(\frac{m^2}{\Lambda_B^2}\Big)-
\frac{1}{2}\sqrt{\Delta}\log\Bigg[\frac{\big(1+\sqrt{\Delta}\big)p^2-2
m^2}{\big(1-\sqrt{\Delta}\big)p^2-2 m^2}\Bigg] + i \pi  \sqrt{\Delta}\Bigg\}.\,\,
\label{eq.app.9.1}
\end{eqnarray}
By using eq.(\ref{eq.app.9}) it is easy to prove that
\begin{eqnarray}
\lim_{p^2\to 0} B_0[p^2,m,m] 
= A_0^{(2)}[m] = 
\frac{i}{(4\pi)^2} \Big[
\frac{2}{4-D}-2-\log\Big(\frac{m^2}{\Lambda_B^2}\Big)\Big]\,.
\label{eq.app.9.2}
\end{eqnarray}

The two-point function in the general case of two different masses, reads
\begin{eqnarray}
&&\!\!\!\!\!\!\!\!\!\!\!
B_0[p^2,m,M] = \frac{i}{(4\pi)^2}\, \Bigg\{\frac{2}{4-D}-\log\Big(\frac{m^2}{\Lambda_B^2}\Big)
-\frac{p^2+M^2-m^2}{2 p^2}\, \log\Big(\frac{M^2}{m^2}\Big)\nonumber\\
&& ~~~~~~~~~~~~~~~~~~~~~~~~~~
-\frac{1}{2}\,\sqrt{\Delta_2}\, \log\Bigg[\frac{\big(1+\sqrt{\Delta_2}\big) p^2-m^2-M^2}
{\big(1-\sqrt{\Delta_2}\big) p^2-m^2-M^2}\Bigg]\Bigg\} \, ,
\label{eq.app.9.4}
\end{eqnarray}
where we have introduced the shorthand notation 
$\Delta_2 = \Big(1-\frac{m^2+M^2}{p^2}\Big)^2-4 \frac{m^2 M^2}{\big(p^2\big)^2}$ .\\
The above expression is valid for $0 < p^2 \leq \big(m-M\big)^2$, 
for $p^2 = \big(m-M\big)^2$, one has $\Delta_2 = 0$. In the kinematical region 
$\big(m-M\big)^2 < p^2 < \big(m+M\big)^2$, the two-point function is given by
\begin{eqnarray}
&&\!\!\!\!\!\!\!\!\!\!\!\!\!\!\!\!\!\!\!\!\!\!\!
B_0[p^2,m,M] = \frac{i}{(4\pi)^2}\, \Bigg\{\frac{2}{4-D}-\log\Big(\frac{m^2}{\Lambda_B^2}\Big)
-\frac{p^2+M^2-m^2}{2 p^2}\, \log\Big(\frac{M^2}{m^2}\Big)\nonumber\\
&& ~~~~~~~
-\sqrt{-\Delta_2}\,\Bigg[\arctan\Bigg(\frac{p^2+m^2-M^2}{p^2\, \sqrt{-\Delta_2}}\Bigg)+
\arctan\Bigg(\frac{p^2-m^2+M^2}{p^2\, \sqrt{-\Delta_2}}\Bigg) \Bigg]\Bigg\} \, .
\label{eq.app.9.5}
\end{eqnarray}
Finally, for $p^2 \geq \big(m+M\big)^2$, one has
\begin{eqnarray}
&&\!\!\!\!\!\!\!\!\!\!\!
B_0[p^2,m,M] = \frac{i}{(4\pi)^2}\, \Bigg\{\frac{2}{4-D}-\log\Big(\frac{m^2}{\Lambda_B^2}\Big)
-\frac{p^2+M^2-m^2}{2 p^2}\, \log\Big(\frac{M^2}{m^2}\Big)\nonumber\\
&& ~~~~~~~~~~~~~~
-\frac{1}{2}\,\sqrt{\Delta_2}\, \log\Bigg[\frac{\big(1+\sqrt{\Delta_2}\big) p^2-m^2-M^2}
{\big(1-\sqrt{\Delta_2}\big) p^2-m^2-M^2}\Bigg]+i\,\pi\,  \sqrt{\Delta_2}\Bigg\} \, .
\label{eq.app.9.6}
\end{eqnarray}

Tensor two point integrals can be reduced to linear combinations of scalar one- and two-point functions.
We consider here the case of a rank one tensor.
\begin{eqnarray}
B^\mu[p^2,m,M] = \int \!\! \frac{d^D q}{(2\pi)^D}\,\, \frac{\mu^{(4-D)}\, q^\mu}{\big(q^2-m^2+i\epsilon\big)
 \big[(q-p)^2-M^2+i\epsilon\big]} = B_1[p^2,m,M]\, p^\mu\,\, ,
\label{eq.app.11}
\end{eqnarray}
where
\begin{eqnarray}
B_1[p^2,m,M] = \frac{p^2+m^2-M^2}{2 p^2}\, B_0[p^2,m,M] + \frac{A_0[M]-A_0[m]}{2 p^2}\,.
\label{eq.app.11.1}
\end{eqnarray}
\end{document}